\documentclass[a4paper]{article}%
\usepackage[T1]{fontenc}
\usepackage{amsmath}
\usepackage{amsfonts}
\usepackage{amssymb}
\usepackage{graphicx}
\usepackage{hyphenat}
\usepackage{hyperref}

\newtheorem{theorem}{Theorem}

\newtheorem{remark}[theorem]{Remark}

\newcommand{\simplebc}{simple boundary conditions}
\newcommand{\classicbc}{classic boundary conditions}
\newcommand{\adaptivebc}{adaptive boundary conditions}
\newcommand{\sbc}{s.b.c.}
\newcommand{\cbc}{c.b.c.}
\newcommand{\abc}{a.b.c.}
\newcommand{\NS}{Navier\hyp{}Stokes}
\newcommand{\tabst}{\rule[-0.5em]{0em}{2em}}
\newcommand{\tabvec}[2]{\begin{pmatrix}#1\\#2\end{pmatrix}}
\newcommand{\domain}{\Omega}
\newcommand{\bubble}{\mathcal{B}}
\newcommand{\truncated}[1]{\check{#1}}
\newcommand{\transient}[1]{\MakeUppercase{#1}}
\newcommand{\shifted}[1]{\tilde{#1}}
\DeclareMathOperator{\erf}{erf}
\DeclareMathOperator{\erfi}{erfi}

\DeclareMathOperator{\Dawson}{D}

\newcommand{\ReN}{\mathrm{Re}}
\newcommand{\vinf}{\boldsymbol{u}_{\infty}}
\newcommand{\vinfM}{u_{\infty}}
\newcommand{\vbody}{\boldsymbol{v}_{\bubble}}
\newcommand{\vbodyM}{v_{\bubble}}
\newcommand{\lv}{\ell_v}
\newcommand{\ie}{\emph{i.e.},\ }

\begin{document}

\title{Artificial boundary conditions for stationary \NS\ flows past bodies in the half-plane}

\author{Christoph Boeckle\thanks{Supported by the Swiss National Science Foundation
(Grant No. 200021-124403).}~\thanks{Corresponding author.}\\{\small Theoretical Physics Department}\\{\small University of Geneva, Switzerland}\\{\small christoph.boeckle@unige.ch}
\and Peter Wittwer\thanks{Supported by the Swiss National Science Foundation (Grant
No. 200021-124403).}\\{\small Theoretical Physics Department}\\{\small University of Geneva, Switzerland}\\{\small peter.wittwer@unige.ch}}
\maketitle

\begin{abstract}
We discuss artificial boundary conditions for stationary \NS\ flows past bodies in the half-plane, for a range of low Reynolds numbers. When truncating the half-plane to a finite domain for numerical purposes, artificial boundaries appear. We present an explicit Dirichlet condition for the velocity at these boundaries in terms of an asymptotic expansion for the solution to the problem. We show a substantial increase in accuracy of the computed values for drag and lift when compared with results for traditional boundary conditions. We also analyze the qualitative behavior of the solutions in terms of the streamlines of the flow. The new boundary conditions are universal in the sense that they depend on a given body only through one constant, which can be determined in a feed-back loop as part of the solution process.

\bigskip

\noindent\textbf{Keywords:}  \NS; exterior domain; fluid-structure interaction; computational fluid dynamics; artificial boundary conditions

\end{abstract}

\tableofcontents

\section{Introduction}

\par We numerically solve the \NS\ equations for the flow past a body moving at constant velocity parallel to the boundary of a half-plane. We are particularly interested in the computation of the hydrodynamic forces acting on the body in the case where the body is small, and the flow is laminar.

\par We first introduce the \emph{viscous length} scale,
\begin{align}
 \lv = \frac{\nu}{\vinfM}=\frac{\nu}{\vbodyM} ~,\label{eq:visclen}
\end{align}
with $\nu$ the dynamic viscosity of the fluid and $\vinfM$ and $\vbodyM$ the magnitude of the velocity field at infinity (as viewed from the moving body) and the magnitude of the body's translation velocity (as viewed from the fluid at rest), respectively. In addition to this dynamic length, there are two geometrical lengths which are important in this problem: the body-center-to-wall (hereafter just \textquotedblleft body-wall\textquotedblright) distance $d$ and the body size $2r$, $d>r$. In terms of the viscous length, we define the Reynolds number $\ReN$ as
\begin{align}
 \ReN = \frac{2r \vbodyM}{\nu} = \frac{2r}{\lv}~.
\end{align}
We shall keep to small but non-negligible values of the Reynolds number throughout this work, \ie $\ReN = 0.5, \ldots, 25$. In this regime, it is expected that the flow remains stationary and that viscous and inertial phenomena have similar importance, \ie the flow is neither creeping nor turbulent.

\par Since we truncate the unbounded domain to finite sub-domains for the numerical treatment, the question of boundary conditions at the resulting artificial boundaries arises. We show that, when compared to traditional methods of \textquotedblleft open\textquotedblright\ boundary conditions (see for example \cite{Griebel.etal-Numericalsimulationin1998}), a significant gain in accuracy in the computed values of drag and lift can be obtained when using the asymptotic expansion for the velocity field constructed in \cite{Boeckle.Wittwer-Asymptoticsofsolutions2011} as Dirichlet boundary conditions. Our method of \emph{\adaptivebc} is universal in the sense that it depends on the boundary conditions at the body surface and the shape of the body only through one constant. We mainly concentrate on drag and lift for the quantitative comparison of different boundary conditions because they are important quantities in engineering and theoretical work alike. Besides increased accuracy, the qualitative behavior of the flow within the computational domain is also improved with our boundary conditions in the sense that the streamlines are not significantly influenced by the artificial boundary, contrary to the cases where traditional boundary conditions are used.

\par We would like to emphasize that the aim of this paper is not to constitute a benchmark for the considered problem nor to achieve the highest possible precision, but rather to highlight the fundamental importance of the choice of boundary conditions when precision and qualitative correctness of the flow patterns are desired alongside a decrease in hardware requirements (especially in memory due to smaller computational domains). In order to make the results easily accessible and useful for applications, we use for the numerical implementation a widely used commercial code targeted for industrial applications, namely COMSOL Multiphysics. See \cite{DiLeonardo.etal-Three-DimensionaltoTwo-Dimensional2011,Urzhumov.Smith-FluidFlowControl2011} for recent research in CFD using COMSOL Multiphysics in low Reynolds numbers regimes.

\par The present work is part of an ongoing project in a bottom-up approach to problems with fluid-structure interaction, starting with the mathematical analysis of the equations and going all the way to numerical applications. Conceptually, we make heavy use of the equivalence between the present problem, which we shall refer to as the \textquotedblleft body-problem\textquotedblright, and a problem without a body, but with a force term of compact support, which we shall refer to as the \textquotedblleft force-problem\textquotedblright. For the mathematical treatment of the Oseen-linearized force-problem, see \cite{Hillairet.Wittwer-vorticityofOseen2008}, for an existence theorem for the nonlinear force-problem see \cite{Hillairet.Wittwer-Existenceofstationary2009}, and for the proof of uniqueness of solutions and the equivalence of the body-problem and the force problem see \cite{Hillairet.Wittwer-Asymptoticdescriptionof2011}. A precise bound on the vorticity for the force-problem was derived in \cite{Boeckle.Wittwer-Decayestimatessolutions2011}, which is the key input for the extraction, in \cite{Boeckle.Wittwer-Asymptoticsofsolutions2011}, of the asymptotic expansion for the velocity field up to second order, which is used in the current work to define the \adaptivebc. For a general introduction to the mathematical method used for the derivation of \adaptivebc\ like the ones presented here, see \cite{Heuveline.Wittwer-ExteriorFlowsat2010}.

\par This work is part of an ongoing effort to achieve higher numerical efficiency in the simulation of exterior problems. Adaptive boundary conditions of the type described here have been used with success for the problem in the full plane, see \cite{Boenisch.etal-Adaptiveboundaryconditions2005}, \cite{Boenisch.etal-Secondorderadaptive2008} and \cite{Latt.etal-Simulatingexteriordomain2006}, and  the three dimensional case is discussed in \cite{Heuveline.Wittwer-AdaptiveBoundaryConditions2010}. In three dimensions, other approaches for the numerical treatment of motions near a wall at low Reynolds numbers have been put into practice. See for example \cite{Chen.etal-Simulationofsingle2003}, where the authors use what we shall refer to as \classicbc, and \cite{Zeng.etal-Wall-inducedforcesrigid2005}, which involves what we shall call \simplebc.
Other types of artificial boundary conditions for incompressible viscous flows in exterior domains have been developed in the past for various cases. For the time-dependent Oseen-linearized \NS\ flow in the full plane, artificial boundary conditions for the normal constraint were proposed in the form of differential equations in \cite{Halpern.Schatzman-Artificialboundaryconditions1989}. For the nonlinear \NS\ equations in the full space, a boundary condition based on the leading asymptotic terms of the solution was obtained in \cite{Nazarov.Specovius-Neugebauer-Nonlinearartificialboundary2003}. For high Reynolds number ($\ReN > 10^7$) flows in aerodynamics (mainly compressible, but including the incompressible limit) artificial boundary conditions based on the Calder\'on-Ryaben'kii method involving difference potentials and pseudo-differential boundary operators has culminated in the work in \cite{Tsynkov-ExternalBoundaryConditions2000}, which includes numerical applications.  Another artificial boundary condition based on a detailed mathematical analysis of bounded domains successively approximating an unbounded one is derived in \cite{Deuring.Kravcmar-ExteriorstationaryNavier-Stokes2004} with error estimates given in \cite{Deuring-Finiteelementerror2009}. A heuristic approach to artificial boundary conditions, the so-called \textquotedblleft no-boundary\textquotedblright\ or \textquotedblleft do nothing\textquotedblright\ boundary conditions, was pioneered in \cite{Malamataris.Papanastasiou-UnsteadyFreeSurface1991,Papanastasiou.etal-newoutflowboundary1992,Heywood.etal-Artificialboundariesand1992}. It consists in extending the governing equations to the artificial boundary (\cite{Griffiths-noboundarycondition1997} showed that this method implicitly imposes the $(p+1)$\textsuperscript{th} derivative of the solution to vanish at a point close to the outflow, where $p$ is the degree of the finite elements), but this approach was applied for domains where only the outflow is an artificial boundary, together with prescribed inflow, and its validity in the case of full exterior domains is unclear. Finally, in \cite{Hasan.etal-outflowboundarycondition2005}, a procedure based on the conservation of mass and vorticity considerations is used to extrapolate the radial velocity at the artificial boundaries for the problem of the flow past a body placed in a channel, with improved behavior of the streamlines at the outlet even in non-stationary cases.

\par Experimental work on single bodies moving slowly parallel to a wall has been a recurring topic for over fifty years, with works such as \cite{Goldman.etal-SlowviscousmotionI1967} and \cite{Ambari.etal-Effectofplane1983}. More recently, in \cite{Takemura.etal-Dragandlift2002} and \cite{Takemura.Magnaudet-transverseforceclean2003}, experiments were reported that show that the transverse force on small spheroidal bubbles moving close to a recipient wall appears to change sign for particular parameters of the flow, the bubble-fluid interface type and small deformations of the bubbles. Two types of bubble-fluid interfaces were studied, so-called \textquotedblleft clean\textquotedblright\ and \textquotedblleft contaminated\textquotedblright\ bubbles, which are modeled mathematically as \textquotedblleft slip\textquotedblright\ and \textquotedblleft noslip\textquotedblright\ boundary conditions, respectively. The authors suggest that the change in sign of the transverse force is due to two competing mechanisms. On the one hand, vorticity generated at the bubble surface is advected and diffused downstream, creating a wake whose symmetry is broken by the wall, resulting in a force pushing the bubble away from the wall. On the other hand, one expects that a contribution to the lift related to the Bernouilli effect would attract the bubble towards the wall. The question is which mechanism dominates. For clean bubbles, which generate less vorticity at the bubble-liquid interface than contaminated bubbles, the sign of the wall-induced lift changes at $\ReN \approx 35$. The bubble shape is known to play a more important role than the Reynolds number for the loss of stability of the paths of rising bubbles \cite{Zenit.Magnaudet-Pathinstabilityof2008}. The greater the aspect ratio between the axis which is perpendicular to the bubble's trajectory and the axis parallel to it, the sooner the instability arises, and this is suggested to be related to an increased vorticity generation on the bubble surface (see \cite{Zenit.Magnaudet-Pathinstabilityof2008}). While this result concerns unsteady flows, we will nevertheless use this finding on the bubble shape, as well as the other experimental insights reported in this paragraph, to guide us in our own numerical work in Section~\ref{sec:val}.

We now introduce the basic mathematical notions for the current work. The \NS\ equations in the time dependent domain $\domain_+\setminus \mathcal{\transient{B}}(t)$, with $\domain_+=\mathbb{R}\times\lbrack 0,\infty)$ and $\mathcal{\transient{B}}(t) = \{\boldsymbol{x}\in\domain_+ \mid (x+\vbodyM t)^{2}+(y-d)^{2}\leq r^{2}\}$ are
\begin{align}
\partial_{t}\boldsymbol{\transient{u}} + \boldsymbol{\transient{u}}\cdot\mathbf{\nabla}\boldsymbol{\transient{u}} + \mathbf{\nabla}\transient{p} - \nu\Delta\boldsymbol{\transient{u}} & = 0~,  \label{eq:nsgeneral} \\
\mathbf{\nabla}\cdot\boldsymbol{\transient{u}} & =0~, \label{eq:incompressibilitygeneral}
\end{align}
where, with $\boldsymbol{x}=(x,y)^{\mathrm{T}}$, $\boldsymbol{\transient{u}=\transient{u}}(\boldsymbol{x},t)$ is the velocity field and $\transient{p}=\transient{p}(\boldsymbol{x},t)$ the pressure field. The boundary conditions at the wall (placed at $y=0$) and at infinity are%
\begin{align}
\left. \boldsymbol{\transient{u}}\right\vert _{y=0} & =0~,  \label{eq:bndwall} \\
\lim_{\boldsymbol{x}\rightarrow\infty}\boldsymbol{\transient{u}} & =0~, 
\label{eq:uinfinity}
\end{align}
whereas on the body we may consider either noslip boundary conditions
\begin{equation*}
\left. \boldsymbol{\transient{u}}\right\vert _{\partial\mathcal{\transient{B}}} = \vbody = (-\vbodyM,0)^{\mathrm{T}}~, 
\end{equation*}
or slip boundary conditions
\begin{align*}
\left. \boldsymbol{\transient{u}\cdot n}\right\vert _{\partial\mathcal{\transient{B}}} & =0~, \\
\left. \boldsymbol{t}\cdot\left[ -\transient{p}\mathbb{I} + \nu \mathcal{D}(\boldsymbol{u})\right] \cdot\boldsymbol{n}\right\vert _{\partial\mathcal{\transient{B}}} & =0~,
\end{align*}
where $\boldsymbol{n}$ and $\boldsymbol{t}$ are respectively the normal and tangential unit vector at the surface of the body, $\mathbb{I}$ is the identity matrix and $\mathcal{D}(\boldsymbol{u})=(\nabla\boldsymbol{u})+(\nabla\boldsymbol{u})^{\mathrm{T}}$.

We are interested in solutions which are stationary when viewed in a reference frame comoving with the body. In terms of the velocity field $\boldsymbol{u}$ relative to the body, we have
\begin{equation}
\boldsymbol{\transient{u}}(\boldsymbol{x},t) = \boldsymbol{u}(\boldsymbol{x} - \vbody t) + \vbody ~,   \label{eq:changeofvariable}
\end{equation}
and $\boldsymbol{u}$ satisfies, in the time-independent domain $\domain = \domain_+\setminus\bubble$, with $\bubble=\{\boldsymbol{x}\in \domain\mid x^{2}+(y-d)^{2}\leq r^{2}\}$, the time-independent equations
\begin{align}
\boldsymbol{u}\cdot\mathbf{\nabla}\boldsymbol{u} + \mathbf{\nabla}p - \nu \Delta\boldsymbol{u} & = 0~,  \label{eq:nssteady} \\
\mathbf{\nabla}\cdot\boldsymbol{u} & = 0~,   \label{eq:incompressibility}
\end{align}
with boundary conditions on the wall and at infinity%
\begin{align}
\left. \boldsymbol{u}\right\vert _{y=0} & =\vinf ~,  \label{eq:bndmvwall} \\
\lim_{|\boldsymbol{x}|\rightarrow\infty}\boldsymbol{u} & = \vinf ~, \label{eq:umvinfinity}
\end{align}
with $\vinf = -\vbody = (\vbodyM,0)^{\mathrm{T}}$. In terms of $\boldsymbol{u}$, the noslip boundary conditions on the body become
\begin{equation}
\left. \boldsymbol{u}\right\vert _{\partial\bubble} = 0~, \label{eq:bndbodynoslip}
\end{equation}
and the slip boundary conditions become
\begin{align}
\left. \boldsymbol{u\cdot n}\right\vert _{\partial\bubble} & = 0~, \label{eq:bndbodyslipa} \\
\left. \boldsymbol{t}\cdot\left[ -p\mathbb{I} + \nu\mathcal{D}(\boldsymbol{u})\right] \cdot\boldsymbol{n}\right\vert _{\partial\bubble} & = 0~. \label{eq:bndbodyslipb}
\end{align}

For the numerical treatment of these equations, we solve (\ref{eq:nssteady}) and (\ref{eq:incompressibility}) in the bounded domain $\truncated{\domain}=\truncated{\domain}_+\setminus\bubble$, where $\truncated{\domain}_+=\{(x,y)\in (-l,l)\times(0,l)\}$ and $l>d+r$ is an arbitrary truncation length. The choice to truncate the domain at equal lengths upstream and downstream is motivated by technical reasons, see \ref{sec:c1algorithm}. Other choices of domain can be considered, but we do not want to indulge here in questions of domain optimization. The truncation introduces artificial boundaries at $x=\pm l$ and $y=l$. The main focus here will be on the choice of boundary conditions on these boundaries, which will be discussed in Section~\ref{sec:bc}. For convenience later on, we note that the drag and the lift on the body are given by $\boldsymbol{F} = (F_{D},F_{L})^{\mathrm{T}}$, where
\begin{equation}
\boldsymbol{F}=\int_{\partial\bubble}(-p\mathbb{I}+\mathcal{D}(\boldsymbol{u}))\boldsymbol{ds}~. \label{eq:force}
\end{equation}

The rest of this paper is organized as follows: in Section~\ref{sec:bc} we present the different boundary conditions which we will implement on the artificial boundaries, in Section~\ref{sec:num} we discuss the numerical aspects of the work and in Section~\ref{sec:val} we present results which validate the \adaptivebc. In Section~\ref{sec:theobehavior}, we show that the theoretical behavior of the flow described in \cite{Hillairet.Wittwer-Asymptoticdescriptionof2011} is numerically verified in the range of simulations we have run. In Section~\ref{sec:forcewall} we present the hydrodynamic forces as a function of the body-wall distance for different sizes of the body. The appendix, finally, contains technical points concerning the \adaptivebc.


\section{\label{sec:bc}Artificial boundaries}

Theoretically, the correct way to treat the edges of a domain $\truncated{\domain}$ which is obtained by a truncation of the half-plane would be to use the solution of the original problem in the half-plane evaluated along those edges  as a Dirichlet boundary condition. Of course, the solution of the original problem is unknown. One must therefore find boundary conditions which represent a good approximation to the solution of the original problem. We shall define and investigate three choices: \emph{simple} boundary conditions (\sbc), \emph{classic} (or open) boundary conditions (\cbc)\ and \emph{adaptive} boundary conditions (\abc), \ie the new scheme which we propose here. More precisely:

\begin{itemize}
\item The \sbc\ simply prescribe $\vinf$ on all the artificial edges.

\item The \cbc\ prescribe $\vinf$ on the upstream vertical boundary in order to fix the inflow, and impose (\ref{eq:bndbodyslipb}) on the remaining artificial boundaries, allowing in- and outflow.

\item The \abc\ use expressions (\ref{eq:asstructu}) and (\ref{eq:asstructv}) which are based on the asymptotic expansion  of the solution of the original problem, to prescribe Dirichlet boundary conditions.
\end{itemize}

The \sbc,\ while a reasonable starting point since they are in particular also used in the construction of weak solutions, are nevertheless problematic, as they do not allow fluid to move through the artificial boundary parallel to the wall, making the problem effectively a channel flow. This impacts flow rate conservation in two problematic ways: first, the velocity of the fluid must increase artificially above and below the body, it cannot be adjusted thanks to fluid \textquotedblleft exiting\textquotedblright\ through the boundary parallel to the wall; second, the flow rate should in fact be lower in the truncated domain in comparison with the flow without a body, however the use of $\vinfM$ at the inlet boundary (\ie the upstream artificial boundary) prescribes the same flow as without a body. See \cite{Zeng.etal-Wall-inducedforcesrigid2005} for an example where such boundary conditions are used in the three-dimensional version of the problem considered here. A recent work using these boundary conditions, albeit for a flow in the full plane around two side-by-side cylinders, is \cite{Vakil.Green-Two-dimensionalsidebysidecircular2011}, where the authors have run simulations in domains with sizes 750 by 500 cylinder radii to ensure that perturbations due to the boundary conditions are small enough.

The \cbc\ are  mixed Dirichlet (pressure) and Neumann (velocity) boundary conditions, and are a standard feature of the COMSOL program. They have been used in problems with artificial boundaries, see for example \cite{Gartling-testproblemoutflow1990} and \cite{Sani.Gresho-Resumeandremarks1994} for the case of an outflow of a channel with a backward-facing step, or again as mentioned before, \cite{Chen.etal-Simulationofsingle2003}, for a three-dimensional implementation of the problem considered here. While the \cbc\ are less restrictive on the flow rate than the \sbc,\ the inlet boundary conditions still prescribe a flow rate which is too large.

We now present our \adaptivebc. As already mentioned before, they are Dirichlet boundary conditions on the velocity, based on the asymptotic expansion for the solution of the problem. Our \adaptivebc\ are given by $\boldsymbol{u}_\ast = (u_\ast,v_\ast)$, with%
\begin{align}
u_\ast(x,y) & = \vinfM \left(1 + \frac{c^\ast_1}{(y/\lv)^{3/2}}\varphi_1(x/y) + \frac{c^\ast_1}{(y/\lv)^2}\varphi_{2,1}(x/y) + \frac{c^\ast_2}{(y/\lv)^2}\varphi_{2,2}(x/y)\right. \notag \\
& \left. - \frac{c^\ast_1}{(y/\lv)^2}\eta_1(\lv x/y^2) - \frac{c^\ast_1}{(y/\lv)^{3}}\eta_2(\lv x/y^2)\right),  \label{eq:asstructu} \\
v_\ast(x,y) & = \vinfM \left( \frac{c^\ast_1}{(y/\lv)^{3/2}}\psi_1(x/y) + \frac{c^\ast_1}{(y/\lv)^2}\psi_{2,1}(x/y) + \frac{c^\ast_2}{(y/\lv)^2}\psi_{2,2}(x/y)\right.  \notag \\
& \left. + \frac{c^\ast_1}{(y/\lv)^3}\omega_1(\lv x/y^2) + \frac{c^\ast_1}{(y/\lv)^4}\omega_2(\lv x/y^2)\right),   \label{eq:asstructv}
\end{align}
where
\begin{align}
\varphi_1(z)     & = -\frac{1}{4\sqrt{\pi}}\frac{r+1-z^{2}+zr+2z}{r^{3}\sqrt{r+1}}~, \label{eq:asphi1} \\
\psi_1(z)        & = -\frac{1}{4\sqrt{\pi}}\frac{r+1-z^{2}-zr-2z}{r^{3}\sqrt{r+1}}~, \label{eq:aspsi1} \\
\varphi_{2,1}(z) & = -\>\frac{1}{\pi}\frac{2z}{r^{4}}~,                                \label{eq:asphi21} \\
\varphi_{2,2}(z) & = \phantom{-} \frac{1}{2\pi}\frac{1-z^{2}}{r^{4}}~,                          \label{eq:asphi22} \\
\psi_{2,1}(z)    & = -\>\frac{1}{\pi}\frac{1-z^{2}}{r^{4}}~,                           \label{eq:aspsi21} \\
\psi_{2,2}(z)    & = -\frac{1}{2\pi }\frac{2z}{r^{4}}~,                              \label{eq:aspsi22}
\end{align}
where%
\begin{equation*}
r=\sqrt{1+z^{2}}~, 
\end{equation*}
and where%
\begin{align}
\eta_1(z) & = \eta_W(z)~, \label{eq:aseta1} \\
\omega_1(z) & = \omega_W(z)~,  \label{eq:asomega1} \\
\eta_2(z) & =\eta_B(z)-2\eta_W(z)-2z\eta_W^{\prime}(z)~, \label{eq:aseta2} \\
\omega_2(z) & =\omega_B(z)-3\omega_W(z)-2z\omega_W^{\prime}(z)~, \label{eq:asomega2}
\end{align}
where \textquotedblleft~$^{\prime}$~\textquotedblright\ in (\ref{eq:aseta2}) and (\ref{eq:asomega2}) denotes the derivative, and
\begin{align}
\eta_W(z) & =-\frac{1}{2\sqrt{\pi z^{3}}}\left\{ 
\begin{array}{cc}
e^{-1/4z}, & z\geq0 \\ 
0, & z<0%
\end{array}
\right. ~,  \label{eq:etaW} \\
\omega_W(z) & =\frac{1}{4\sqrt{\pi z^{5}}}%
\begin{cases}
(1-2z)e^{-1/4z}, & z\geq0 \\ 
0, & z<0%
\end{cases}
~,  \label{eq:omegaW} \\
\eta_B(z) & =-\frac{1}{4\pi z^3} \left( 2z+\sqrt{\pi|z|}(1-2z) \left(e^{-1/4z}-\Dawson{(1/2|z|^{1/2})}\right) \right)~,  \label{eq:etaB} \\
\omega_B(z) & =\frac{1}{8\pi z^4} \left( 2z(1-4z)+\sqrt{\pi|z|}(1-6z) \left(e^{-1/4z}-\Dawson{(1/2|z|^{1/2})}\right) \right)~, \label{eq:omegaB}
\end{align}
where
\begin{align*}
\Dawson{(z)} = \begin{cases}
e^{-z^2}\erfi{z}~, & z\geq 0\\
e^{z^2}\erf{z}~, & z< 0
\end{cases}~,
\end{align*}
is the Dawson function, $\erf$ is the Gauss error function
\begin{align*}
 \erf{z} = \frac{2}{\sqrt{\pi}}\int_0^z e^{-\zeta^2}d\zeta~,
\end{align*}
and $\erfi(z) = -i \erf(iz)$ is the imaginary error function.

For a derivation of the \adaptivebc, see \ref{sec:explicitasfun}. In particular, note that we set $c^\ast_2=0$ throughout the present work. Therefore, the \adaptivebc\ only reference the characteristics of the body (size, position, shape, boundary conditions at the interface) through the constant $c^\ast_1\in\mathbb{R}$. This constant should be equal to a constant based on the solution in the original unbounded domain, defined in \cite{Boeckle.Wittwer-Asymptoticsofsolutions2011} and represented there by $c_1$. A good approximation can be determined as part of the solution process, for example by the algorithm described in the next paragraph.
\par In essence, our algorithm used to determine $c^\ast_1$ searches for the root of the function
\begin{equation}
  g(x) = \frac{1}{n_1}\int_{\truncated{\domain}}\nabla\cdot (\mathrm{T}(\boldsymbol{u},p) \boldsymbol{V})d\omega - x~, \label{eq:gdef}
\end{equation}
where
\begin{equation*}
\mathrm{T}(\boldsymbol{u},p)=-\boldsymbol{u}\otimes\boldsymbol{u} + \nu \mathcal{D}(\boldsymbol{u})-p\mathbb{I}
\end{equation*}
is calculated with $\boldsymbol{u}$ and $p$ computed from the numerical solution obtained with the \abc\ used with $c^\ast_1$ such that $g(c^\ast_1)=0$ and where $\otimes$ represents the dyadic product (\ie $(\boldsymbol{a}\otimes\boldsymbol{b})_{ij} = a_ib_j$), and 
\begin{equation*}
\boldsymbol{V} = (\sqrt{y}\chi_{\truncated{\domain}}(x,y),0)^{\mathrm{T}}~, 
\end{equation*}
with $\chi_{\truncated{\domain}}(x,y)=\chi_{x}(x)\cdot\chi_{y}(y)$ a user-defined differentiable cut-off function that cuts a channel perpendicular to the wall centered on the body and two strips, one adjacent to the wall and one adjacent to the artificial boundary parallel to the wall. The scalar $n_1$ in (\ref{eq:gdef}), given by
\begin{align*}
n_1 & = \int_1^{\infty}\chi_y(l/z)\left(\lv^{3/2} \frac{\varphi_1(z)-\psi_1(z)}{z} + 2\lv^2 \frac{\varphi_{2,1}(z)}{\sqrt{zl}}\right)dz\\
& -\int_{\lv/l}^{\infty}\chi_y(\sqrt{\lv l/z})\left( \lv^2\frac{\eta_1(z)}{(\lv lz^3)^{1/4}} + \lv^3\frac{\eta_2(z)-\eta_2(-z)}{\left(\lv^3 l^3 z\right)^{1/4}}\right)dz~,
\end{align*}
is independent of the solution. Note that, $c^\ast_1$ and $n_1$ depend on $l$ and that for $l\to\infty$ the function $g(x)$ vanishes for $x=c_1$ (\ie the exact constant associated to the solution of the original problem), see \ref{sec:c1algorithm} for more details.
\par The root-finding algorithm, in our case Brent's method (see \cite{Brent-algorithmwithguaranteed1971}), operates on a sequence of simulations, refining $c^\ast_1$ at each step. Since Brent's method starts with the bisection method, we must first bracket the root. We do this by imposing the heuristically derived values $c_{1}^{(i),\ast}=10\ReN (i-1)d$ with $i=1,2,\ldots$, so that for the first run the simulation coincides with the one for the \sbc\ As soon as the root is bracketed, we begin Brent's method until the desired tolerance is achieved.

\bigskip

We finally discuss the regime of applicability of our \adaptivebc\ with regard to the size $l$ of the computational domain. There are three conditions that have to be respected. First, $l_{\mathrm{min}}$ has to be large enough in order for the asymptotic expansion to be indeed evaluated in its domain of validity, \ie $y$ should be large enough for the ratio $\lv x/y^2$ to be small. Thus, when $x\sim y$ (see \ref{sec:explicitasfun} for a motivation why $x$ large is also a valid region), we must have $l_{\mathrm{min}} =y \sim C_1\, \lv$. Second, the artificial boundary must be beyond any standing eddies which are not taken into account by our asymptotic expansion. On the basis of measurements by Gerrard (see \cite[p.~362]{Gerrard-WakesofCylindrical1978}), standing eddies typically extend up to three times the body diameter at $\ReN \approx 40$, so that we choose $l_{\mathrm{min}} =x\sim C_2 r $. Thirdly, the wake should have interacted with the wall by the time it reaches the artificial boundary, because the asymptotic expansion is made under that assumption. Because the wake scales as $\lv x/y^2$ and the body is at $y=d$, we must have $l_{\mathrm{min}}=x \sim C_3\,\lv^{-1} d^2$. Collecting the requirements, one gets
\begin{align}
 l_{\mathrm{min}} \sim \max\{C_1\,\lv ,\ C_2\,r,\ C_3\,\lv^{-1}\,d^2\}~.\label{eq:minimaldomainsize}
\end{align}
Of course, these requirements are qualitative only, since we have no way to estimate the constants $C_1$, $C_2$ and $C_3$. They do however explain the observation that the \adaptivebc\ are particularly useful in an intermediate range of parameters, to be specified in what follows.


\section{\label{sec:num}Numerics}

To numerically solve (\ref{eq:nssteady})--(\ref{eq:bndmvwall}) with all the different choices of boundary conditions, we use COMSOL Multiphysics 3.5a, controlled through a Matlab 2009a script. The linear system is solved using a direct linear solver, PARDISO, and the nonlinear solver is a damped Newton fixed point solver. The mesh elements are a mix of triangles and quadrilaterals of Lagrange type, of degree two for the velocity and degree one for the pressure. For all the simulations presented below, a workstation with 36~GB~RAM and a 12-core processor was used. The algorithms mentioned above are features of the COMSOL program, and we refer to the product documentation for additional information.


\section{\label{sec:val}Validation}

The first set of numerical tests is used to validate the \adaptivebc. For the sake of convenience, we have set $2r=1$ for all simulations of this validation run, so that the Reynolds number is given by the inverse of the viscous length. The body is at a fixed distance $d=1$ from the wall. We show the effect of the domain size for all choices of boundary conditions on the computation of the drag and lift.

\subsection{Mesh}

\begin{figure}[!h]
 \begin{center}
  \includegraphics[width=0.7\textwidth]{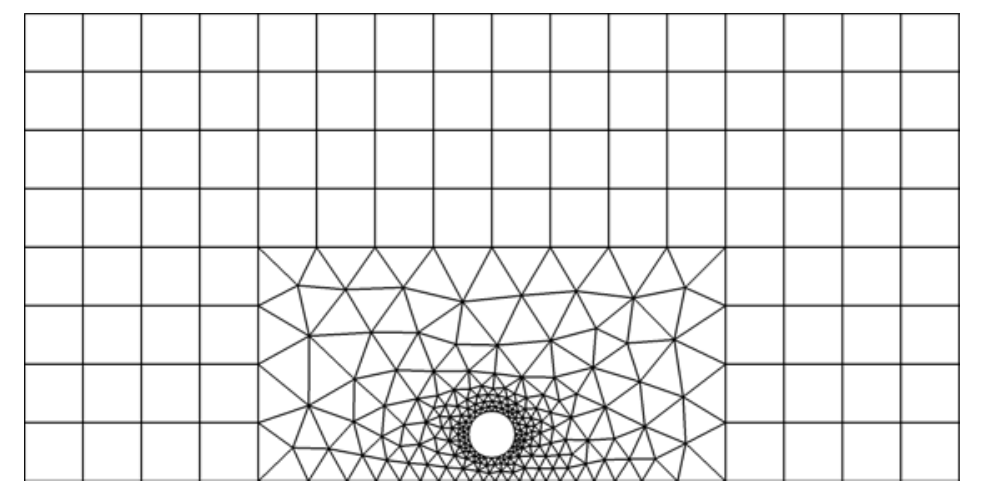}
 \end{center}
 \caption{\label{fig:mesh}The coarsest mesh ($h=1.25$) on the smallest domain ($l=10$) used in the simulations of Section~\ref{sec:val}.}
\end{figure}

We now present our method for generating meshes for the domain $\truncated{\domain}$. Figure~\ref{fig:mesh} represents the coarsest and smallest mesh used, from which all others are constructed. The smallest mesh will have a domain with truncation length $l=10$ in the same units as the body size.

We adopt the same philosophy as \cite{Hasan.etal-outflowboundarycondition2005}: in order to minimize possible effects from one mesh to another due to element placement, we define a small rectangular zone (of size $10$ by $5$) around the body meshed with a constant node distance $h$ on the top, left and right boundaries, and using triangles placed according to an advancing front algorithm provided by COMSOL. Since the body size is fixed, this mesh will be the same for all simulations. The rest of the domain is paved with strictly identical square elements with sides of length $h$. Increasing the domain size then consists in adding supplementary square elements of the same size, guaranteeing that the meshes always have the same structure regardless of domain size. The base mesh for $l=10$ has $h_{0}=1.25$, 367 triangles and 96 squares, for a total of 2718 degrees of freedom. A step of mesh refinement simply consists in dividing all mesh cell lengths by a factor $m_{r}=2$ effectively dividing each element into four identical smaller ones. Thus the recurrence relation $h_{i}=h_{i-1}/m_{r}$.

We compute simulations with mesh sizes $h_{0}$ to $h_{3}$ on domains with truncation lengths $l = 10, 20, \ldots, 90$. All figures in this article representing the velocity fields are obtained from simulations computed on meshes with elements of size $h_{3}$. Since the drag and the lift are given by integrals, we take advantage of the fact that each of our successively refined meshes is simply a subdivision of the preceding one and use a Richardson extrapolation scheme (see \cite{Richardson-ApproximateArithmeticalSolution1911}) to accelerate convergence. We base our scheme on the hypothesis that the error of the drag and the lift (\ref{eq:force}) is of the form
\begin{equation}
I_{h_{n}}=I+C_{2}h_{n}^{2}+C_{3}h_{n}^{3}+C_{4}h_{n}^{4}+\mathcal{O}(h_{n}^{5})~, \label{eq:errorTheory}
\end{equation}
with $C_{i}$ unknown constants. This yields the extrapolation formula 
\begin{align}
I_{R} & = \frac{m_{r}^{9}\cdot I_{h_{3}} - (m_{r}^{7}+m_{r}^{6}+m_{r}^{5})\cdot I_{h_{2}} + (m_{r}^{4}+m_{r}^{3}+m_{r}^{2})\cdot I_{h_{1}} - I_{h_{0}}} {m_{r}^{9}-(m_{r}^{7}+m_{r}^{6}+m_{r}^{5})+(m_{r}^{4}+m_{r}^{3}+m_{r}^{2}) -1} + \mathcal{O}(h_{0}^{5})  \notag \\
& = \frac{512\cdot I_{h_{3}}-224\cdot I_{h_{2}}+28\cdot I_{h_{1}}-I_{h_{0}}}{315} + \mathcal{O}(h_{0}^{5})~,   \label{eq:richardsonExtrap}
\end{align}
for the Richardson extrapolate $I_{R}$ of $I$. The error terms in (\ref{eq:errorTheory}) start at $h_n^2$ because the Lagrange elements are of degree two for the velocity and one for the pressure. Note that if our Ansatz (\ref{eq:errorTheory}) should miss intermediary terms of order $h^{n}$, $n>2$, then they are not eliminated by the scheme, but only somewhat damped. Thus, the extrapolation scheme will always diminish the magnitude of the error, even though not necessarily to order $\mathcal{O}(h_{0}^{5})$.

\subsection{\label{sec:valresults}Results}

\subsubsection{\label{sec:valresultsqual}Qualitative aspects}

\begin{figure}[!h]
 \begin{center}
  \includegraphics[width=\textwidth]{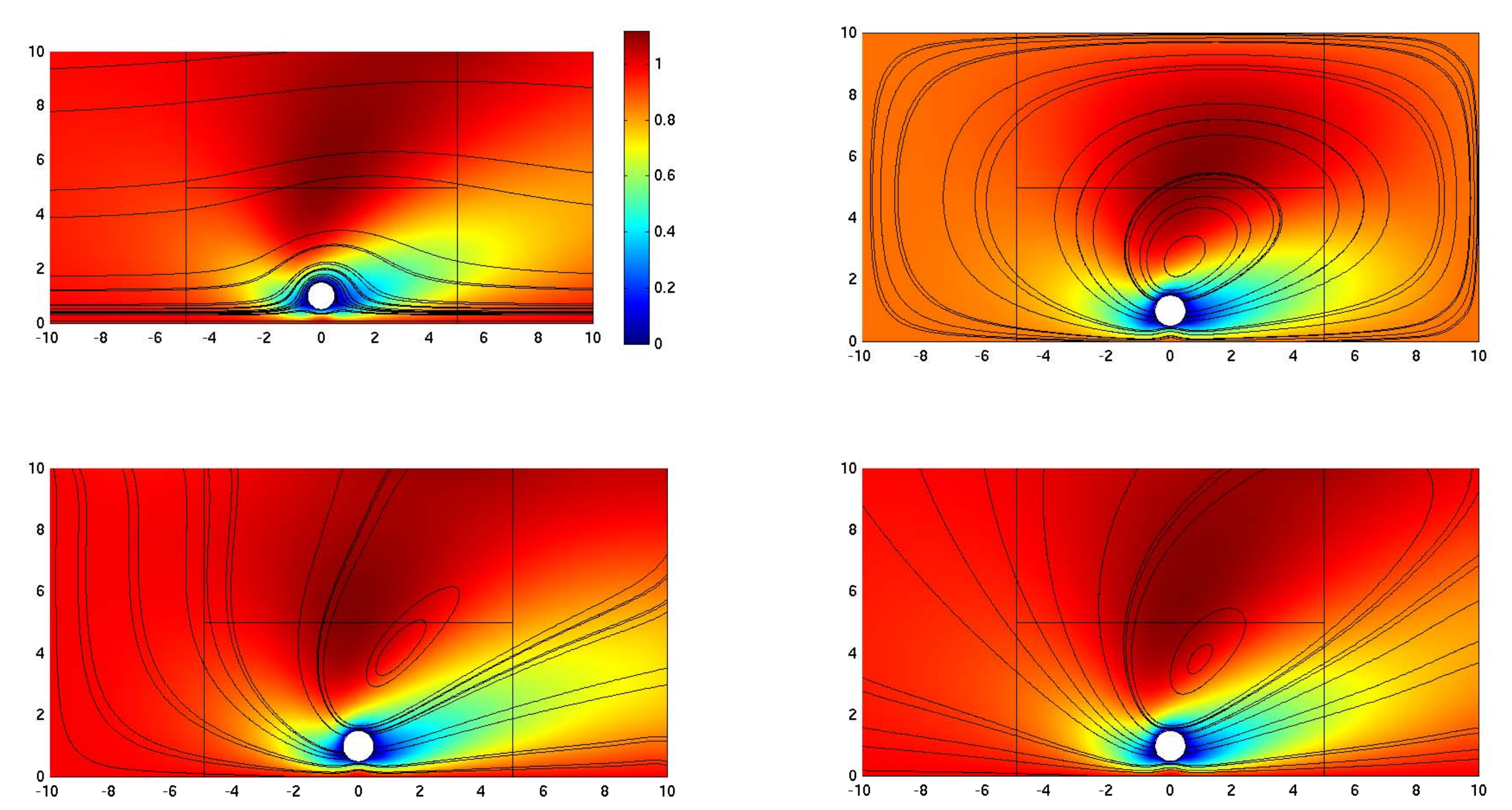}
 \end{center}
 \caption{\label{fig:streamBC}Streamlines for a domain with truncation length $l=10$ and Reynolds number $\ReN=1$, for a body with  noslip boundary conditions. The colored map represents the velocity norm. Streamlines (top left). Streamlines as seen from the body: \sbc\ (top right), \cbc\ (bottom left) and \abc\ (bottom right).}
\end{figure}

\begin{figure}[!h]
 \begin{center}
  \includegraphics[width=\textwidth]{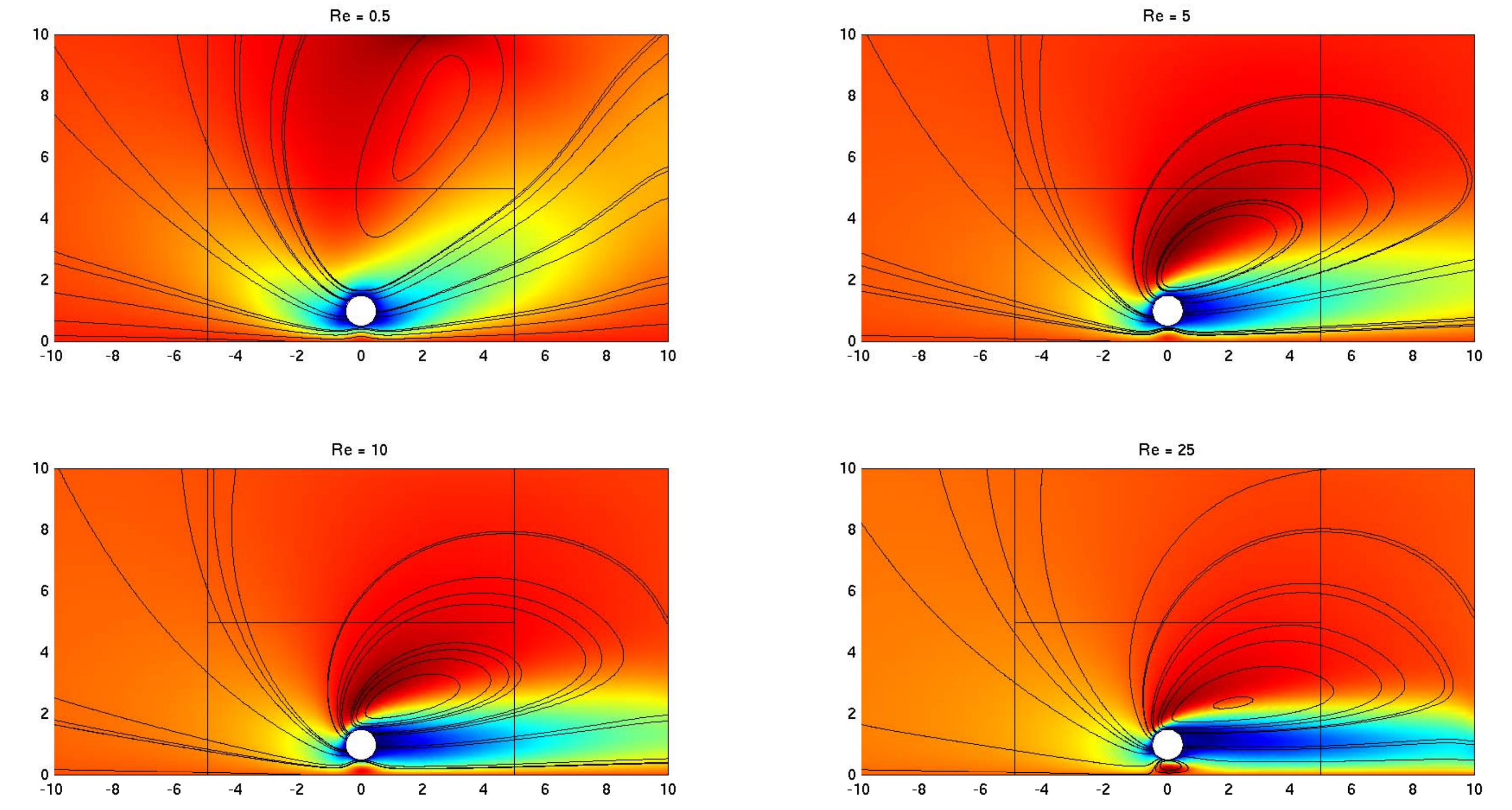}
 \end{center}
 \caption{\label{fig:streamRe}Streamlines for a domain with truncation length $l=10$ with \adaptivebc, and for a body with noslip boundary conditions. The colored map represents the velocity norm. $\ReN = 0.5,5,10,25$ (top to bottom, left to right).}
\end{figure}

In a first step, we show that the choice of boundary conditions has an important impact on the quantities being calculated. To showcase this, we represent in Figure~\ref{fig:streamBC} the streamlines for the velocity field as seen from the body, \ie the velocity field from which the constant flow $\boldsymbol{u}_{\infty}$ has been subtracted. In the case of the \simplebc\ (\sbc), we observe an artificial backflow (moving clockwise on our figure). A very important proportion of the computational domain is thus used to compute this non-physical flow. In the case of the \classicbc\ (\cbc), this backflow is not present, but the flow is still significantly influenced by the artificial boundaries, as can be seen by looking at the streamlines. The \adaptivebc\ (\abc)\ are the only ones that yield qualitatively satisfying flows all the way up to the boundary, and the streamlines exhibit almost no distortion near the boundary. Figure~\ref{fig:streamRe} presents the streamlines for Reynolds numbers $\ReN = 0.5,5,10,25$.

We now discuss the qualitative behavior of the flow computed with the \abc\ in more detail. The case of $\ReN=25$ shows signs of artificial behavior at the right-hand boundary, which could be due to the fact that the asymptotic expansion ignores the actual position of the body. More precisely, the \abc\ assume that the wake has already interacted with the boundary before the position of the artificial boundary. The wake is asymptotically governed by the first order in the $\lv x/y^2$-scaling of the asymptotic expansion, and larger Reynolds numbers therefore yield narrower wakes which interact with the wall farther downstream, see (\ref{eq:minimaldomainsize}). In order to simulate higher $\ReN$ flow, one would have to bring the body closer to the wall, which would change the actual problem being solved, or use a larger domain, which was outside our computational power. This limitation to $\ReN \leq 25$ is not so stringent, because in the absence of any wall, the flow behind a circular body becomes unsteady at $\ReN \approx 30$, see \cite[p.~150]{Dyke-Perturbationmethodsin1975}, and we do not expect that the situation be radically different in our problem.

\begin{figure}[!h]
 \begin{center}
  \includegraphics[width=\textwidth]{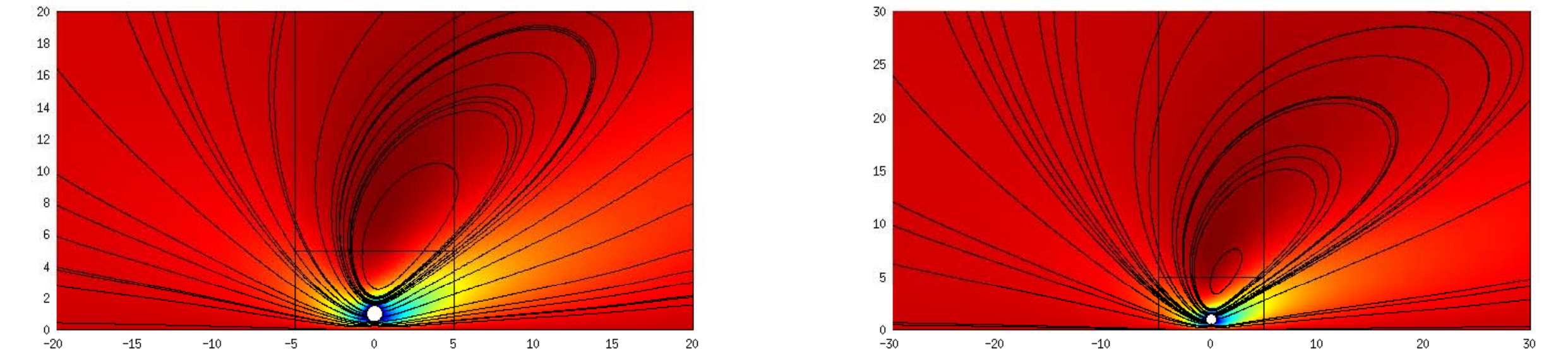}
 \end{center}
 \caption{\label{fig:largerRe0.5}At $\ReN=0.5$, a larger domain diminishes the distortions in the streamlines.}
\end{figure}

The case of $\ReN=0.5$, also shows distortions in the streamlines, especially downstream. Since $\ReN = \lv^{-1}$ in the validation run, flows with Reynolds numbers smaller than unity need a large domain too, see (\ref{eq:minimaldomainsize}). Indeed, since the wake scales like $\lv x/y^2$, a low Reynolds number actually delays the distance after which the solution may be represented by the asymptotic expansion (which is obtained for $y\to\infty$, \ie $\lv x/y^2\to 0$), so that the \abc\ should be used in conjunction with larger numerical domains. As can be seen in Figure~\ref{fig:largerRe0.5}, increasing the domain size does indeed improve the behavior of the streamlines. This case is different from the large Reynolds number case, since the wake does interact with the wall, but the flow region which is represented is too small compared to the viscous length for the asymptotic expansion to be accurate (the flow would probably be more accurately represented under the Stokes approximation). It is nevertheless possible to simulate flows for lower Reynolds numbers, but one must respect particular scaling conditions. This procedure will be presented in Section~\ref{sec:theobehavior} where we will investigate a sequence of simulations with progressively smaller bodies.

\subsubsection{\label{sec:valresultsquant}Quantitative analysis}

\begin{figure}[!h]
 \begin{center}
  \includegraphics[width=\textwidth]{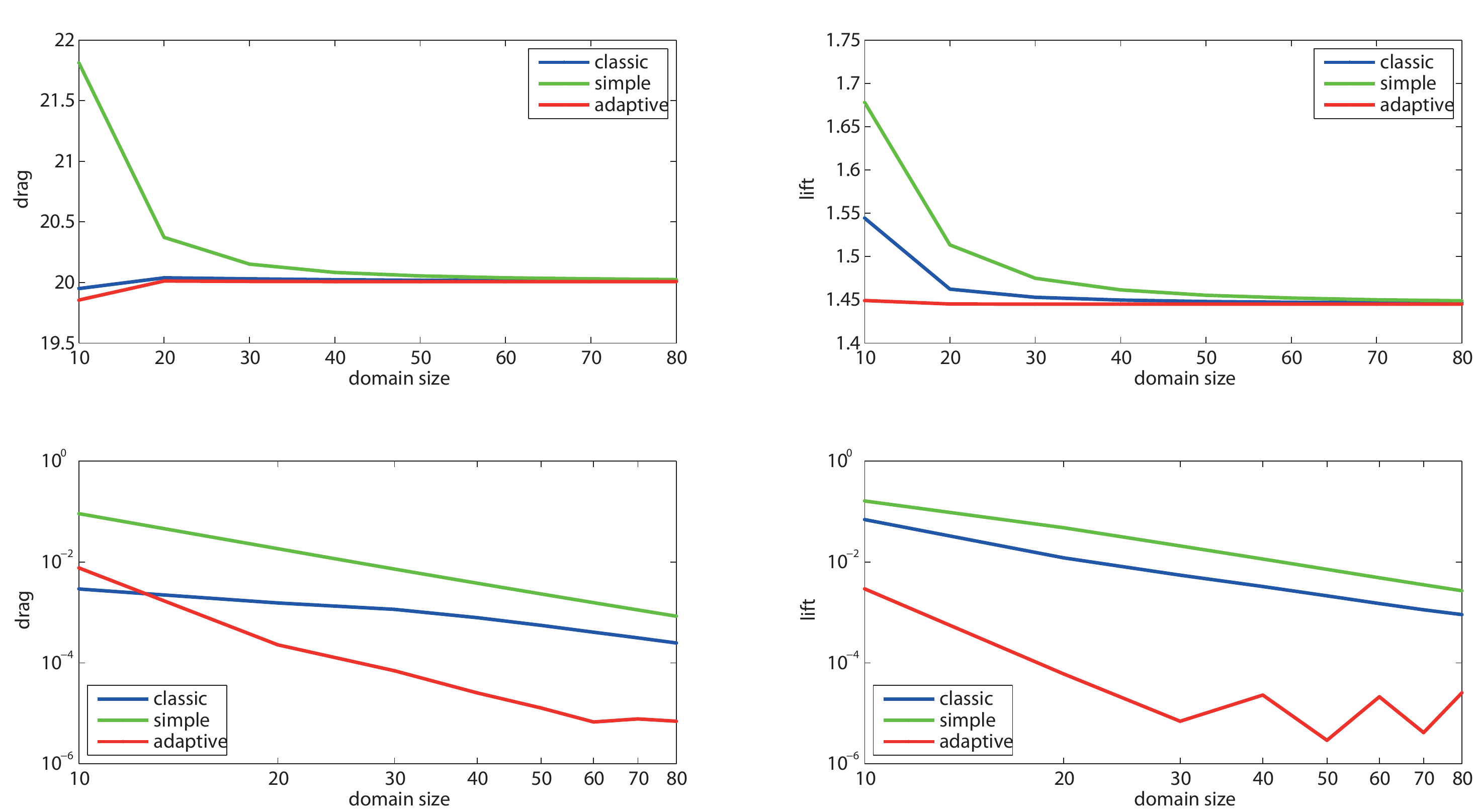}
 \end{center}
 \caption{\label{fig:fN05}$\ReN=0.5$ simulations, for a body with noslip boundary conditions. Top: Drag (left) and lift (right) as a function of domain size for the three boundary conditions. Bottom: relative error on drag (left) and lift (right) as a function of domain size (log-log scales).}
\end{figure}

\begin{figure}[!h]
 \begin{center}
  \includegraphics[width=\textwidth]{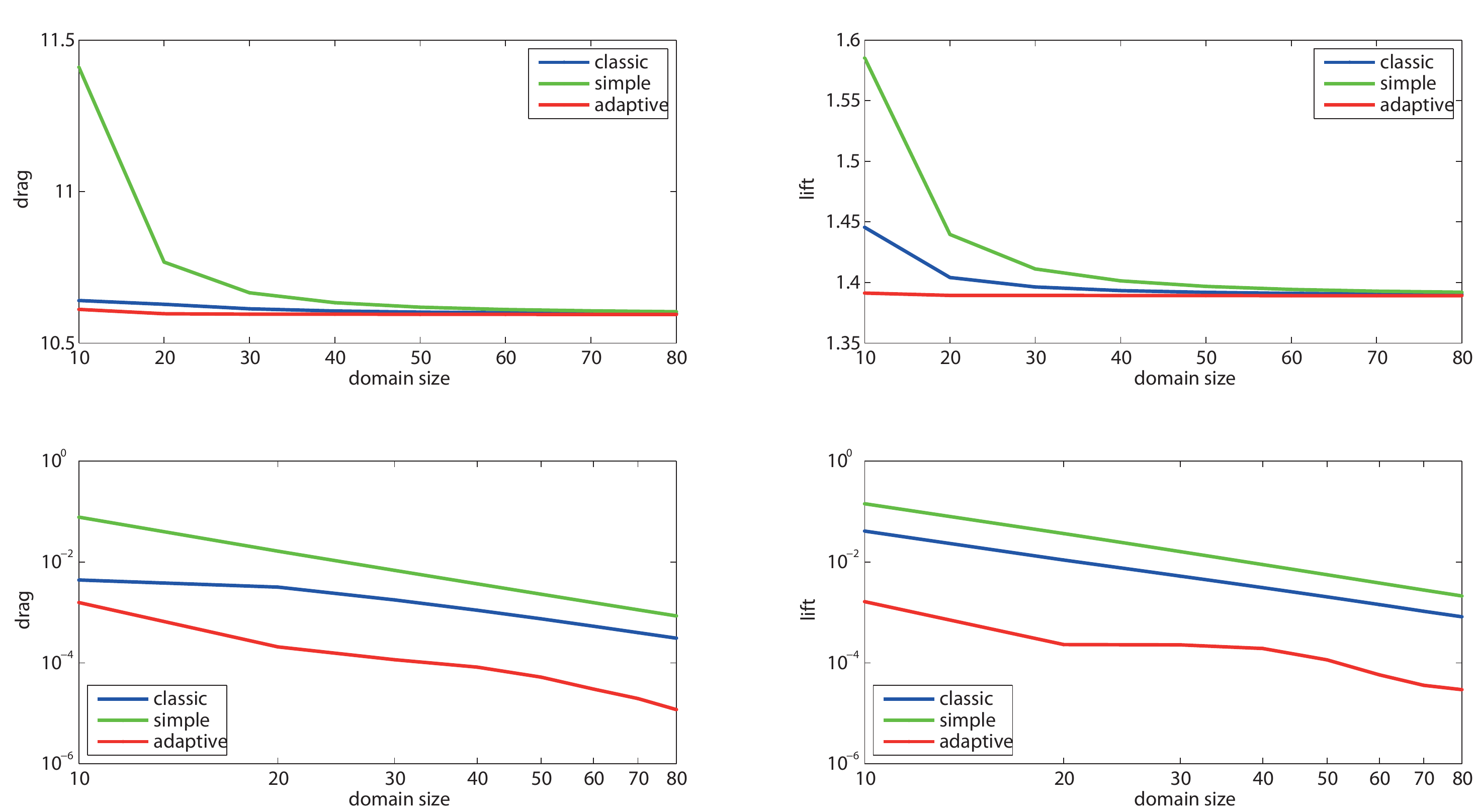}
 \end{center}
 \caption{\label{fig:fN1}$\ReN=1$ simulations, for a body with noslip boundary conditions. Top: Drag (left) and lift (right) as a function of domain size for the three boundary conditions. Bottom: relative error on drag (left) and lift (right) as a function of domain size (log-log scales).}
\end{figure}

\begin{figure}[!h]
 \begin{center}
  \includegraphics[width=\textwidth]{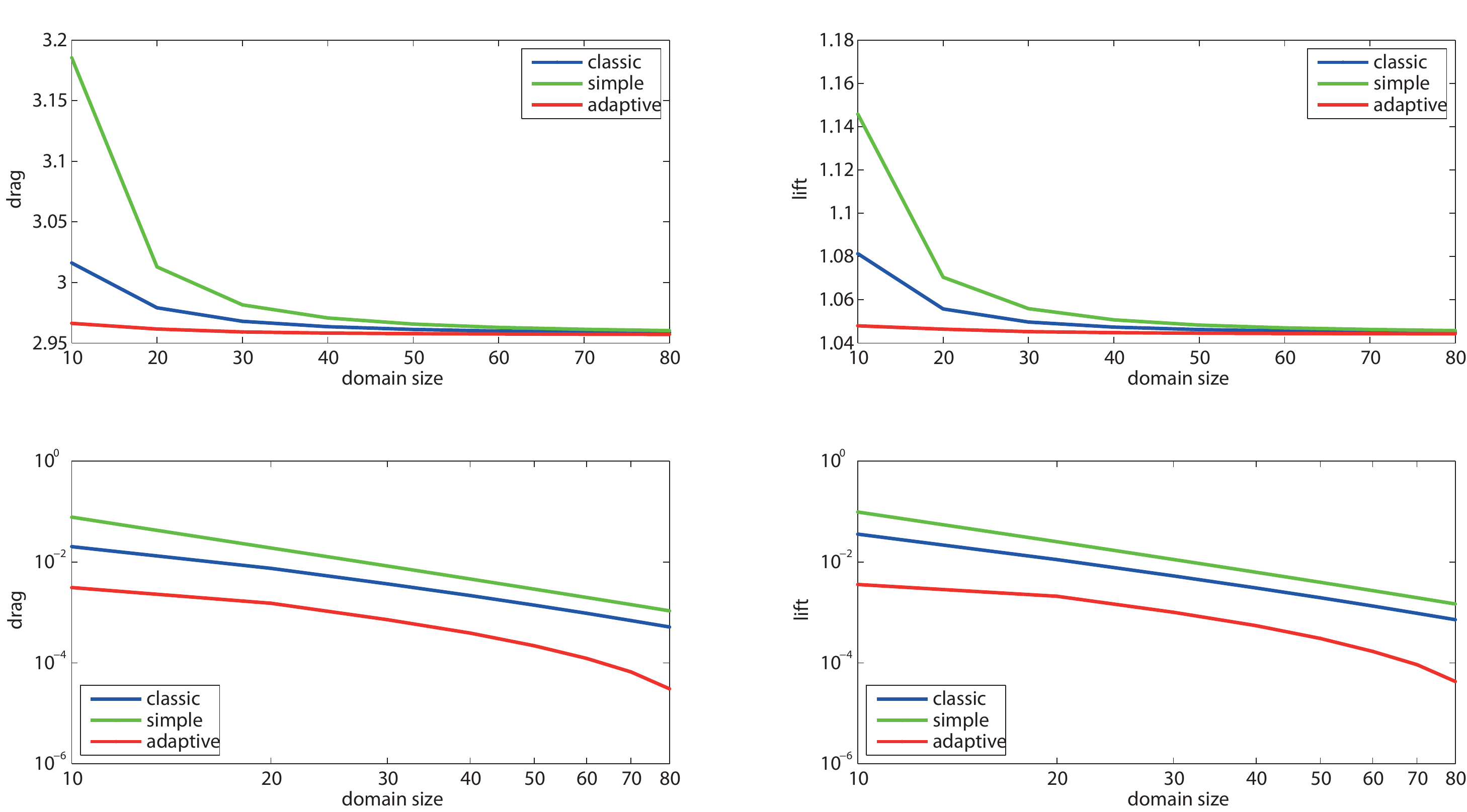}
 \end{center}
 \caption{\label{fig:fN5}$\ReN=5$ simulations, for a body with noslip boundary conditions. Top: Drag (left) and lift (right) as a function of domain size for the three boundary conditions. Bottom: relative error on drag (left) and lift (right) as a function of domain size (log-log scales).}
\end{figure}

\begin{figure}[!h]
 \begin{center}
  \includegraphics[width=\textwidth]{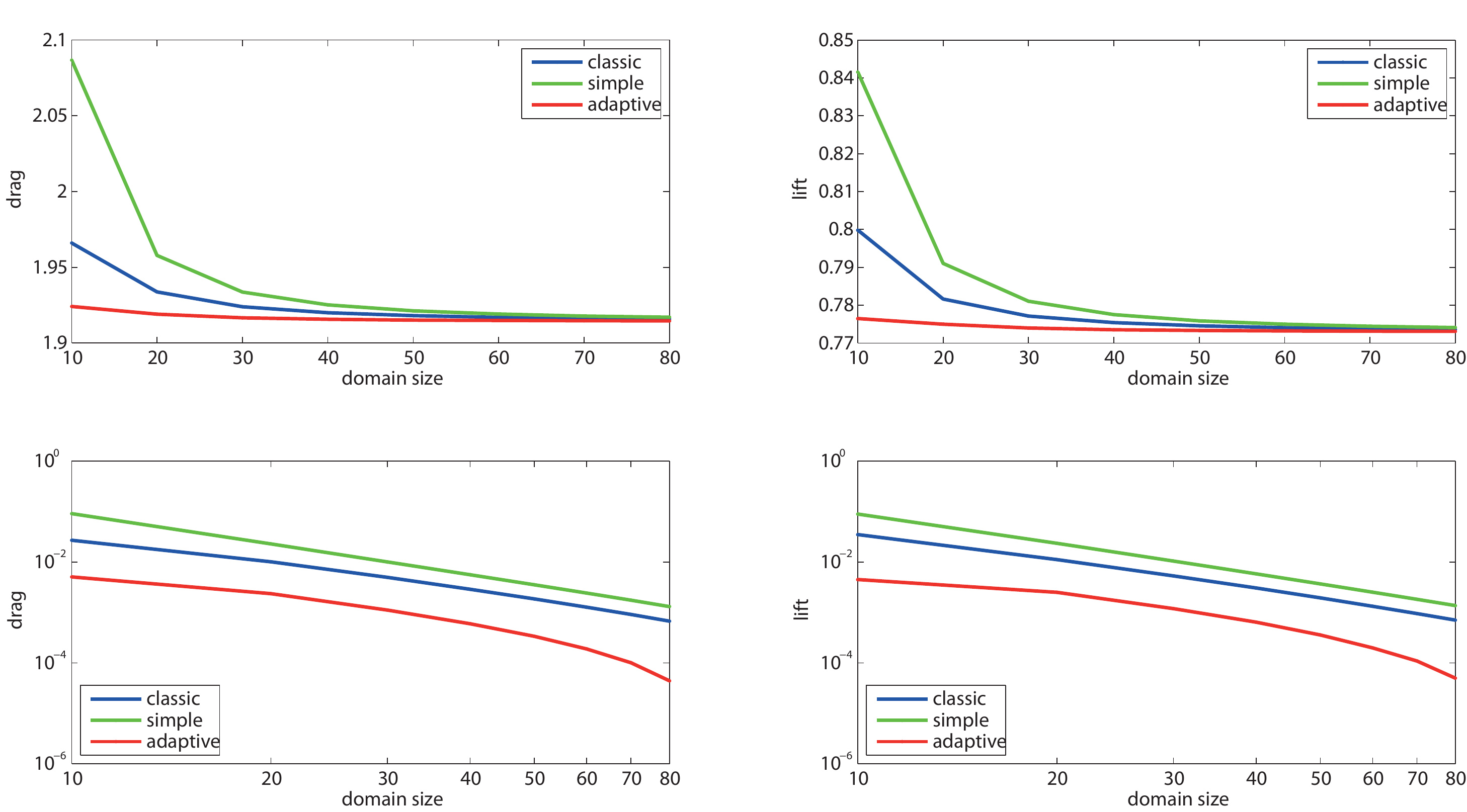}
 \end{center} 
 \caption{\label{fig:fN10}$\ReN=10$ simulations, for a body with noslip boundary conditions. Top: Drag (left) and lift (right) as a function of domain size for the three boundary conditions. Bottom: relative error on drag (left) and lift (right) as a function of domain size (log-log scales).}
\end{figure}

\begin{figure}[!h]
 \begin{center}
  \includegraphics[width=\textwidth]{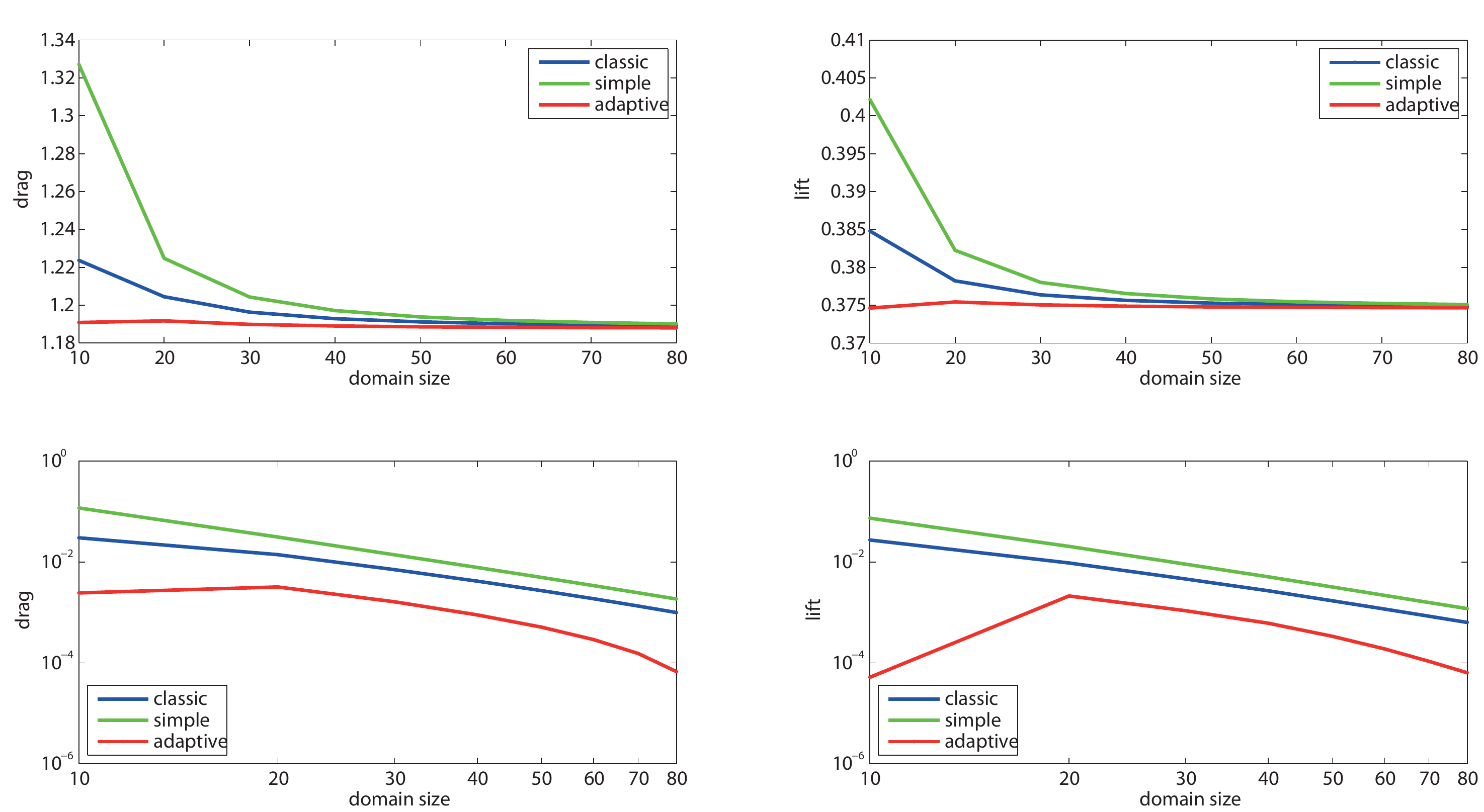}
 \end{center}
 \caption{\label{fig:fN25}$\ReN=25$ simulations, for a body with noslip boundary conditions. Top: Drag (left) and lift (right) as a function of domain size for the three boundary conditions. Bottom: relative error on drag (left) and lift (right) as a function of domain size (log-log scales).}
\end{figure}

We present in Figures \ref{fig:fN05}--\ref{fig:fN25} the drag and lift for a sequence of simulations with Reynolds numbers $\ReN=0.5,1,5,10,25$, performed with the three boundary conditions. As the simulations with \adaptivebc\ overall present the least variation with domain size, we use the largest simulation ($l=90$) feasible on our workstation using these boundary conditions as a reference for the computation of relative errors.

\begin{figure}[!h]
 \begin{center}
  \includegraphics[width=\textwidth]{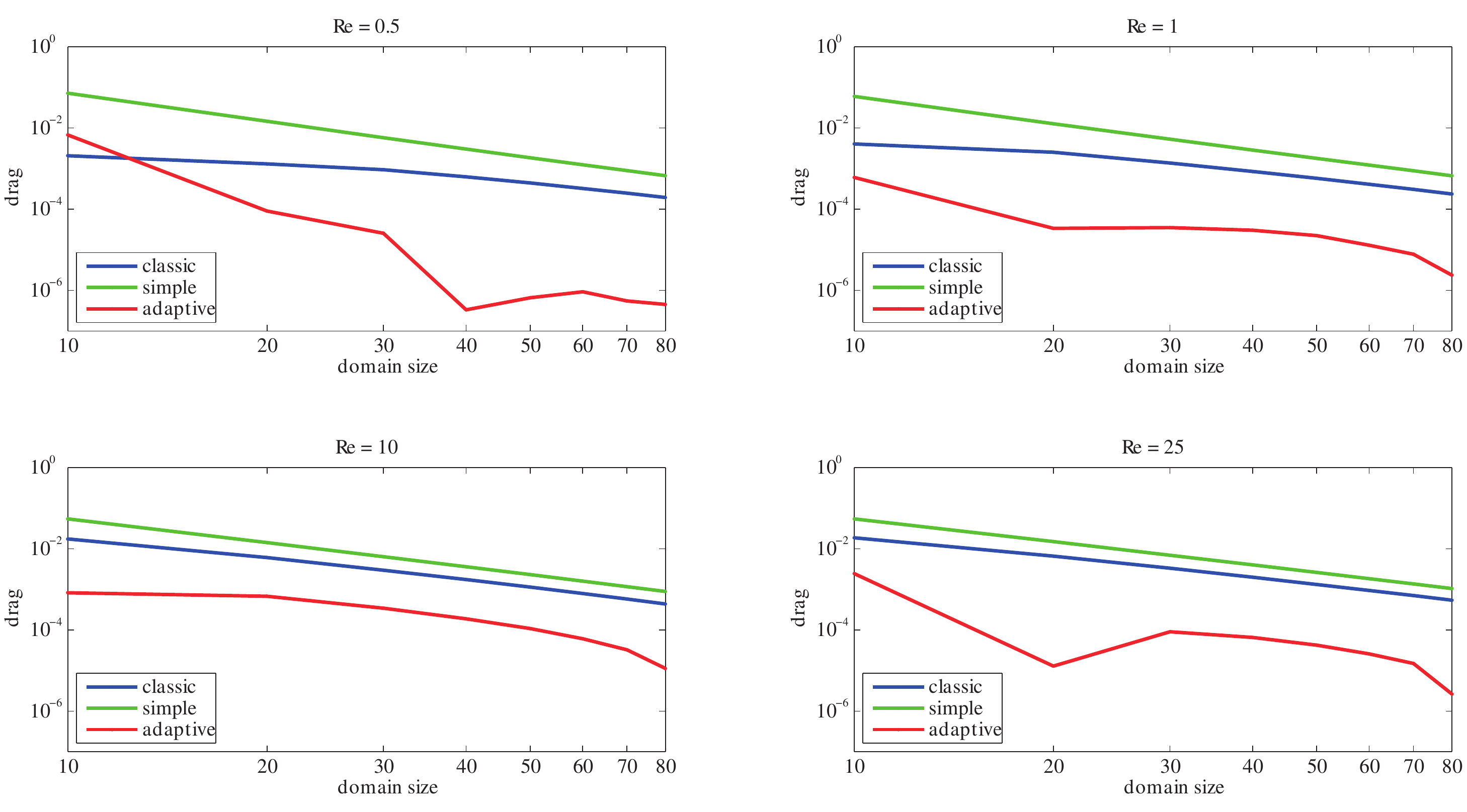}
 \end{center}
 \caption{\label{fig:eDS}Simulations for a body with slip boundary conditions. Relative error for the drag as a function of domain size (log-log scales), for $\ReN = 0.5, 1, 10$ and $25$.}
\end{figure}

\begin{figure}[!h]
 \begin{center}
  \includegraphics[width=\textwidth]{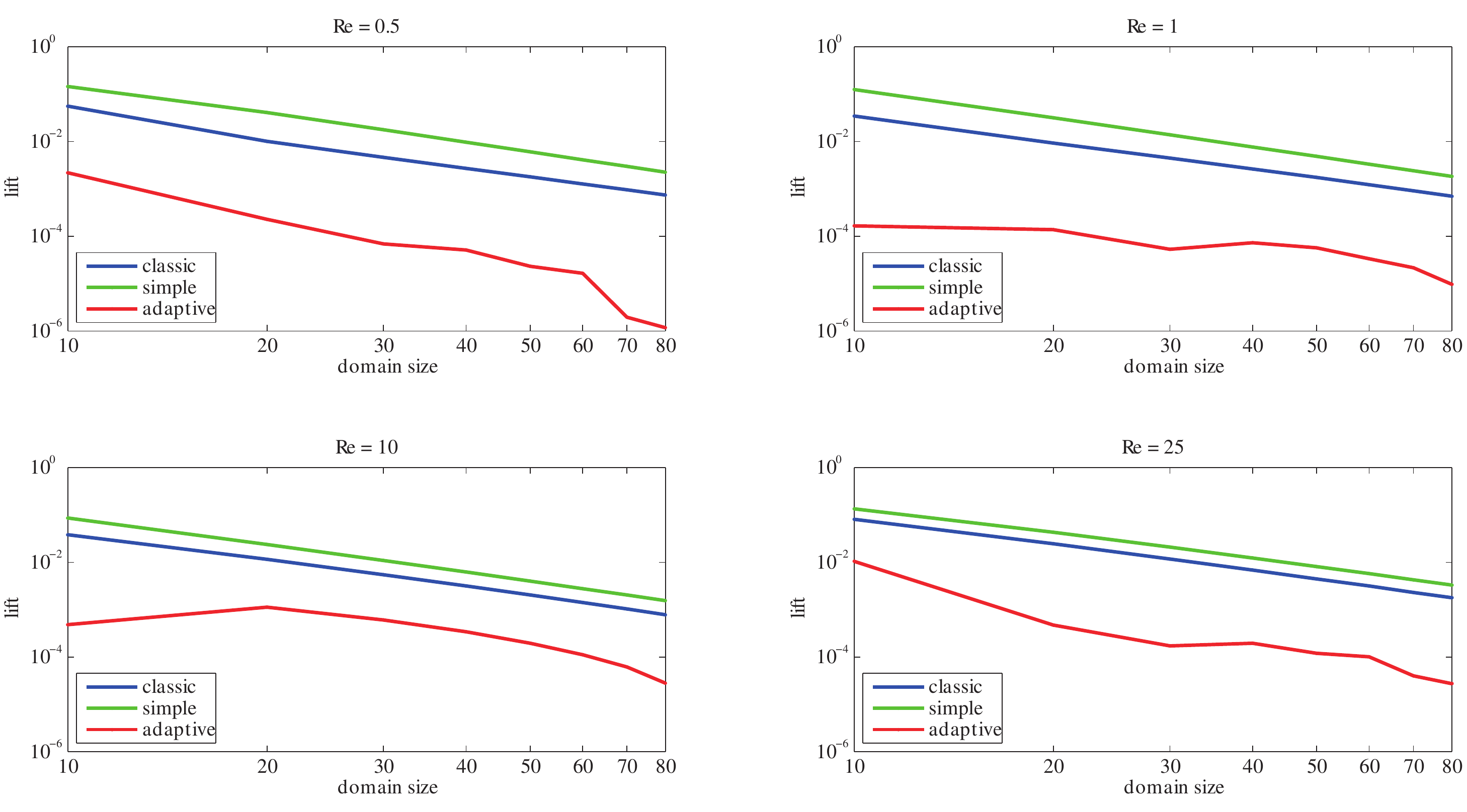}
 \end{center}
 \caption{\label{fig:eLS}Simulations for a body with slip boundary conditions. Relative error for the lift as a function of domain size (log-log scales), for $\ReN = 0.5, 1, 10$ and $25$.}
\end{figure}

In Figures~\ref{fig:eDS}--\ref{fig:eLS} we summarize the relative errors for the drag and lift obtained for a body with slip boundary conditions.

In Tables~\ref{tab:forcesnoslip}--\ref{tab:forcesslip} we present the extrapolated values of the drag and lift for circular objects with noslip and slip boundary conditions at Reynolds numbers $\ReN = 0.5, 1, 5, 10, 25$. The extrapolation was done using formula (\ref{eq:richardsonExtrap}) on the values obtained from simulations with mesh sizes $h_0$ to $h_3$, for the largest domain ($l=90$).

\begin{table}
\begin{center}
 \begin{tabular}{ll|ccccc}
    \multicolumn{7}{c}{\textquotedblleft Noslip body\textquotedblright} \\
    \tabst & & $\ReN=0.5$  & $\ReN=1$ & $\ReN=5$ & $\ReN=10$ & $\ReN=25$ \\
    \hline
    \hline
	     & \sbc \tabst & $20.020$ & $10.601$ & $2.9595$ & $1.9164$ & $1.1896$ \\
   drag: & \cbc \tabst & $20.011$ & $10.597$ & $2.9582$ & $1.9155$ & $1.1888$ \\
	     & \abc \tabst & $20.007$ & $10.594$ & $2.9571$ & $1.9145$ & $1.1880$ \\
	\hline
	     & \sbc \tabst & $1.4480$ & $1.3912$ & $1.0454$ & $0.77384$ & $0.37497$ \\
   lift: & \cbc \tabst & $1.4460$ & $1.3898$ & $1.0448$ & $0.77344$ & $0.37481$ \\
	     & \abc \tabst & $1.4450$ & $1.3890$ & $1.0442$ & $0.77304$ & $0.37464$
 \end{tabular}
\end{center}
\caption{\label{tab:forcesnoslip}Extrapolated values for drag and lift computed with the largest domain ($l=90$) for a body with noslip boundary conditions.}
\end{table}

\begin{table}
\begin{center}
 \begin{tabular}{ll|ccccc}
    \multicolumn{7}{c}{\textquotedblleft Slip body\textquotedblright} \\
    \tabst &  & $\ReN=0.5$ & $\ReN=1$ & $\ReN=5$ & $\ReN=10$ & $\ReN=25$ \\
    \hline
    \hline
         & \sbc \tabst & $14.505$ & $7.6636$ & $2.0929$ & $1.3023$ & $0.71367$ \\
   drag: & \cbc \tabst & $14.500$ & $7.6611$ & $2.0922$ & $1.3018$ & $0.71338$ \\
         & \abc \tabst & $14.497$ & $7.6597$ & $2.0916$ & $1.3014$ & $0.71307$ \\
    \hline
         & \sbc \tabst & $0.88168$ & $0.84854$ & $0.59285$ & $0.37608$ & $0.10188$ \\
   lift: & \cbc \tabst & $0.88065$ & $0.84779$ & $0.59250$ & $0.37585$ & $0.10176$ \\
         & \abc \tabst & $0.88013$ & $0.84732$ & $0.59219$ & $0.37563$ & $0.10162$ 
 \end{tabular}
\end{center}
\caption{\label{tab:forcesslip}Extrapolated values for drag and lift computed with the largest domain ($l=90$) for a body with slip boundary conditions.}
\end{table}

We observe that, for both boundary conditions on the body, the \simplebc\ overestimate the values of the forces compared to the \classicbc\, which in turn overestimate the drag and lift compared to \adaptivebc, except for the drag in the case $\ReN=0.5$ on the smallest domain. This was to be expected, since both the \sbc\ and \cbc\ impose a flow rate only appropriate in \emph{absence} of a body. In a bounded domain, this flow rate is higher than when there is a body, leading to an inevitable overestimation in the hydrodynamic forces. However, the modification of the flow rate due to the body decays as the inverse of the square root of the size of the domain, so that in the limit of a complete half-plane, the difference vanishes. This is different from the case of a body in the whole space, where the presence of abody imposes a non-vanishing modification to the flow rate (see \cite{Boenisch.etal-Adaptiveboundaryconditions2005} for the asymptotic expansion for the velocity field in this case). Nevertheless, in the case of truncated domains we show that the difference in flow rate implied by the various boundary conditions has a non-negligible effect.

\subsection{\label{sec:shape}Body shape}

\par
\begin{figure}[!h]
 \begin{center}
  \includegraphics[width=0.7\textwidth]{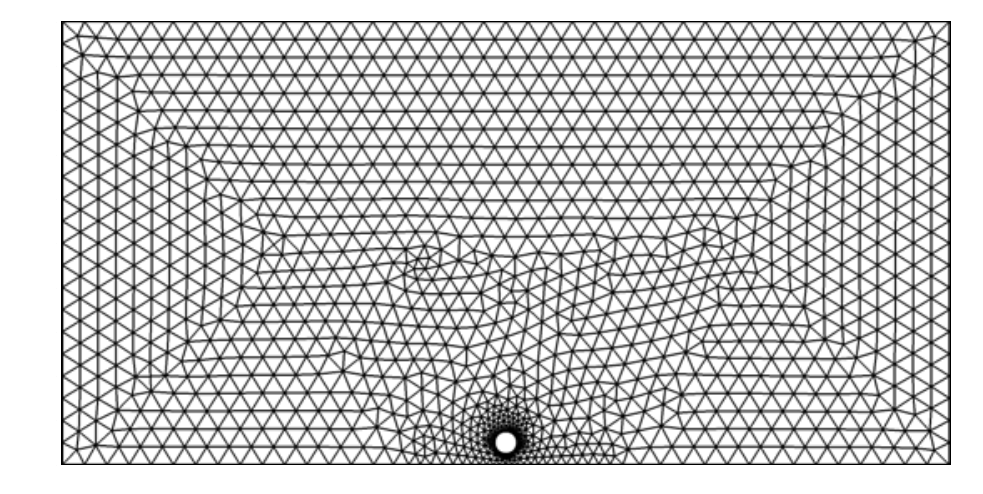}
 \end{center}
 \caption{\label{fig:meshOther}Mesh composed of triangles obtained from COMSOL's advancing front algorithm on a domain with truncation length $l=20$, a circular body with $r=0.5$ and $d=1.0$.}
\end{figure}

\begin{figure}[!h]
 \begin{center}
  \includegraphics[width=\textwidth]{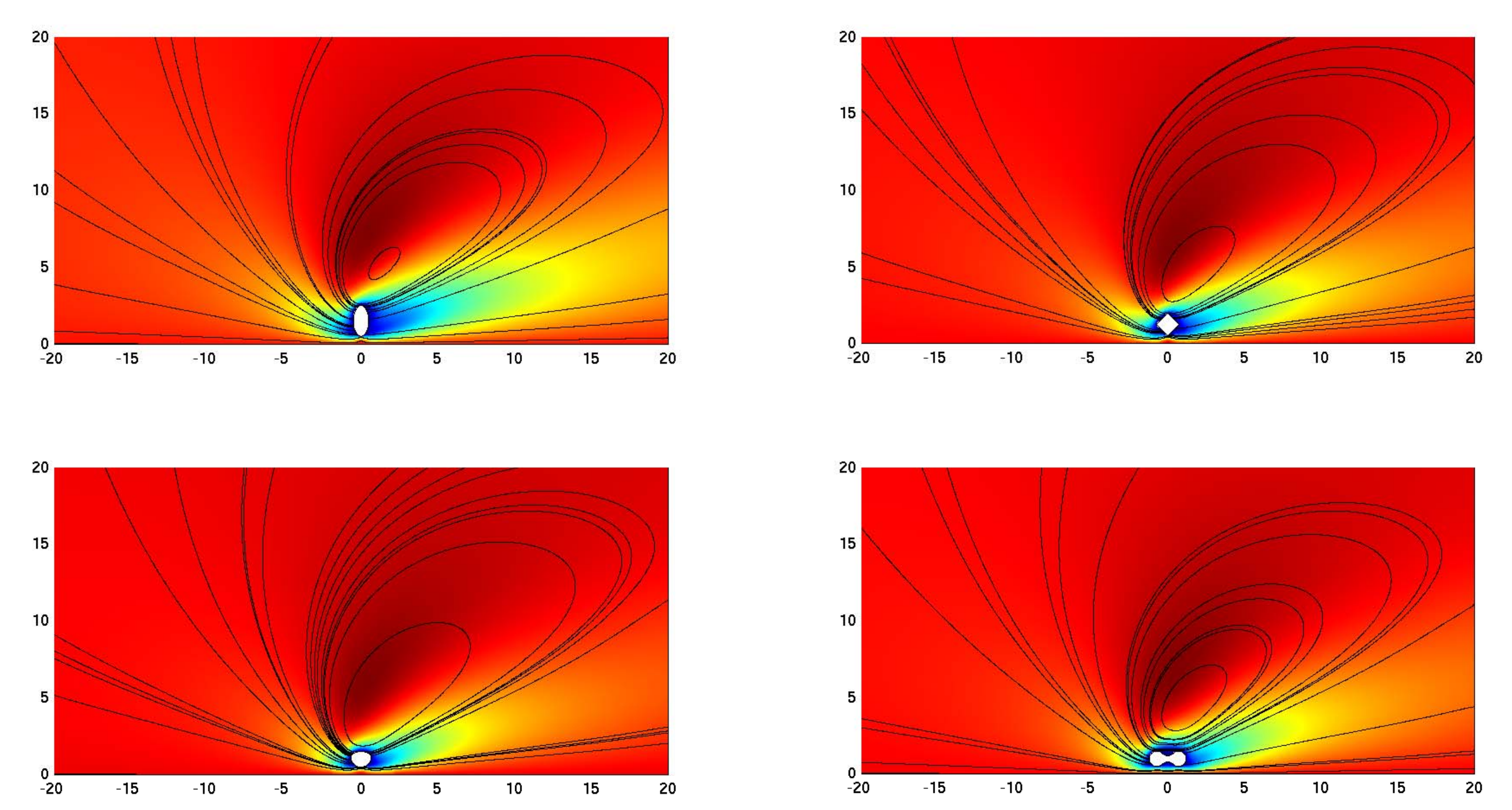}
 \end{center}
 \caption{\label{fig:varBodies}Top left: ellipse; top right: square; bottom left: \textquotedblleft marmite\textquotedblright; bottom right: \textquotedblleft bone\textquotedblright.}
\end{figure}

 We next demonstrate the range of applicability of the \abc\ All simulations were performed using a triple refinement of the mesh shown in Figure~\ref{fig:meshOther}, obtained according to the advancing front algorithm provided by COMSOL (maximum triangle side is $1.0$). We recall that one step of refinement is defined as the division of all triangles into four smaller ones of equal size. The characteristic body size is defined by the projection of the shape on the $y$-axis. We present four shapes in Figure~\ref{fig:varBodies}: an ellipse with its large axis perpendicular to the flow of length $1.0$ and height $2.0$, a square of side $\sqrt{9/8}$, a \textquotedblleft marmite\textquotedblright\ composed of half a circle and half an ellipse whose longer axis is twice the smaller one, of total height $1.0$, and a \textquotedblleft bone\textquotedblright\ of length $1+\sqrt{2}$ and height $1.0$. These bodies serve to illustrate that the \abc\ work well for shapes with concavities and asymmetries, as well as non-smooth boundaries, even though the theoretical developments are expressly derived for smooth bodies.

\par
\begin{figure}[!h]
 \begin{center}
  \vspace*{1cm} 
  \includegraphics[width=\textwidth]{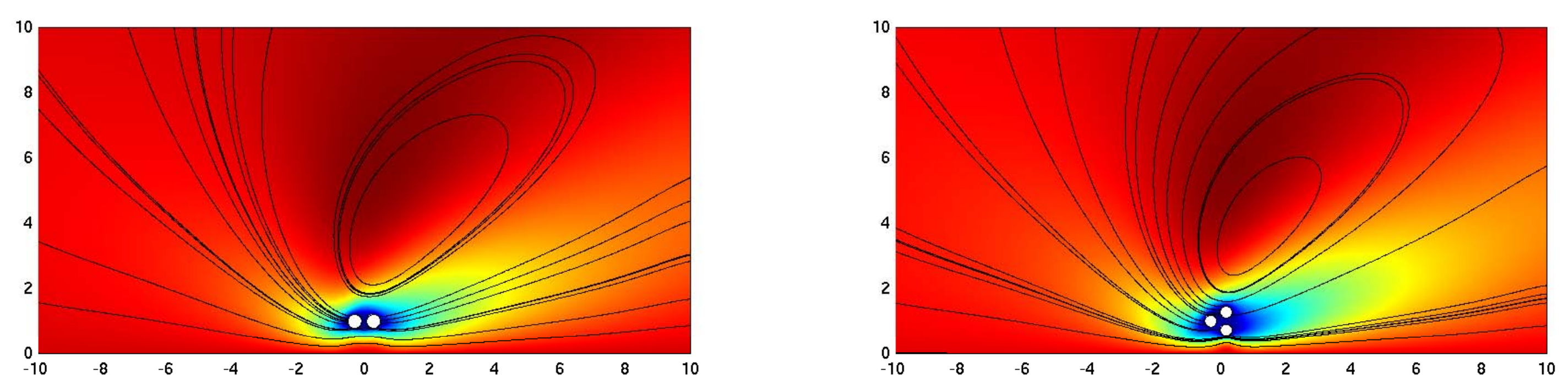}
 \end{center}
 \caption{\label{fig:compBodies}Collections of two and three circles.}
\end{figure}

 It is also possible to simulate collections of bodies. We tested arrangements of two and three circles, contained in a circular area of radius $r=1.0$, see Figure~\ref{fig:compBodies}. These two simulations were done on a smaller domain, with $l=10$ (the mesh was generated in the same way as for the other shapes).

\section{\label{sec:theobehavior}Conformance to expected theoretical behavior}

\par A result of \cite[Section~2.3]{Hillairet.Wittwer-Asymptoticdescriptionof2011} states that solutions of (\ref{eq:nssteady}) and (\ref{eq:incompressibility}) tend to zero when the obstacle size tends to zero. This theoretical result may seem obvious at first, but because of the Stokes paradox this statement requires a proof. Since the FEM resolution scheme used by COMSOL is based on a Galerkin method, it is natural to test this result. In the present work we examine the relative velocity, so that the result shows that for vanishing body size, the flow tends to $\vinf$.

\par Having validated our scheme, we exclusively use the \adaptivebc from now on. In what follows, the viscous length is fixed for all simulations at $\lv=1$, all the bodies have noslip boundary conditions, and their sizes are chosen as $2r = 0.005, 0.01, 0.05, 0.1, 0.5, 1, 2$. For reasons of computational tractability and because of the requirement of placing the artificial boundary far enough to ensure that they are evaluated in the asymptotic physical regime, we choose the truncation length of the domain to be $l=20$ for all simulations. This size is already quite impressive compared to the size of the bodies. Since we work with one fixed domain size only, the meshes are for simplicity chosen to be triangular pavements of the fluid domains, as given by COMSOL and similar to that of Figure~\ref{fig:meshOther}. We however go as far as quadruple refinement (leading to approximately 3 million d.o.f.)\ and then apply an appropriate Richardson extrapolation, similar to (\ref{eq:richardsonExtrap}). Meshing difficulties arise with smaller bodies, in the sense that the mesh elements become smaller near the body, thus posing a problem with regard to memory or mesh quality (ratio of the smallest to the largest element).

\par Following the proof in \cite{Hillairet.Wittwer-Asymptoticdescriptionof2011}, we wish to provide evidence that the norm of the velocity gradient integrated over the fluid domain vanishes as the body disappears. The norm of the velocity gradient is given by
\begin{align}
 \Vert \nabla \boldsymbol{u} \Vert = \sqrt{(\partial_x u)^2 + (\partial_y u)^2 + (\partial_x v)^2 + (\partial_y v)^2}~. \label{eq:gradunorm}
\end{align}
For the sake of comparison, we also compute the norm of the velocity field as seen by the body, \ie
\begin{align}
 \Vert \boldsymbol{u} - \vinf \Vert = \sqrt{(u-\vinfM)^2+v^2}~,\label{eq:urelnorm}
\end{align}
and both norms are integrated over the whole computational domain and normalized by its surface.

\par
\begin{figure}[!h]
 \begin{center}
  \includegraphics[width=0.75\textwidth]{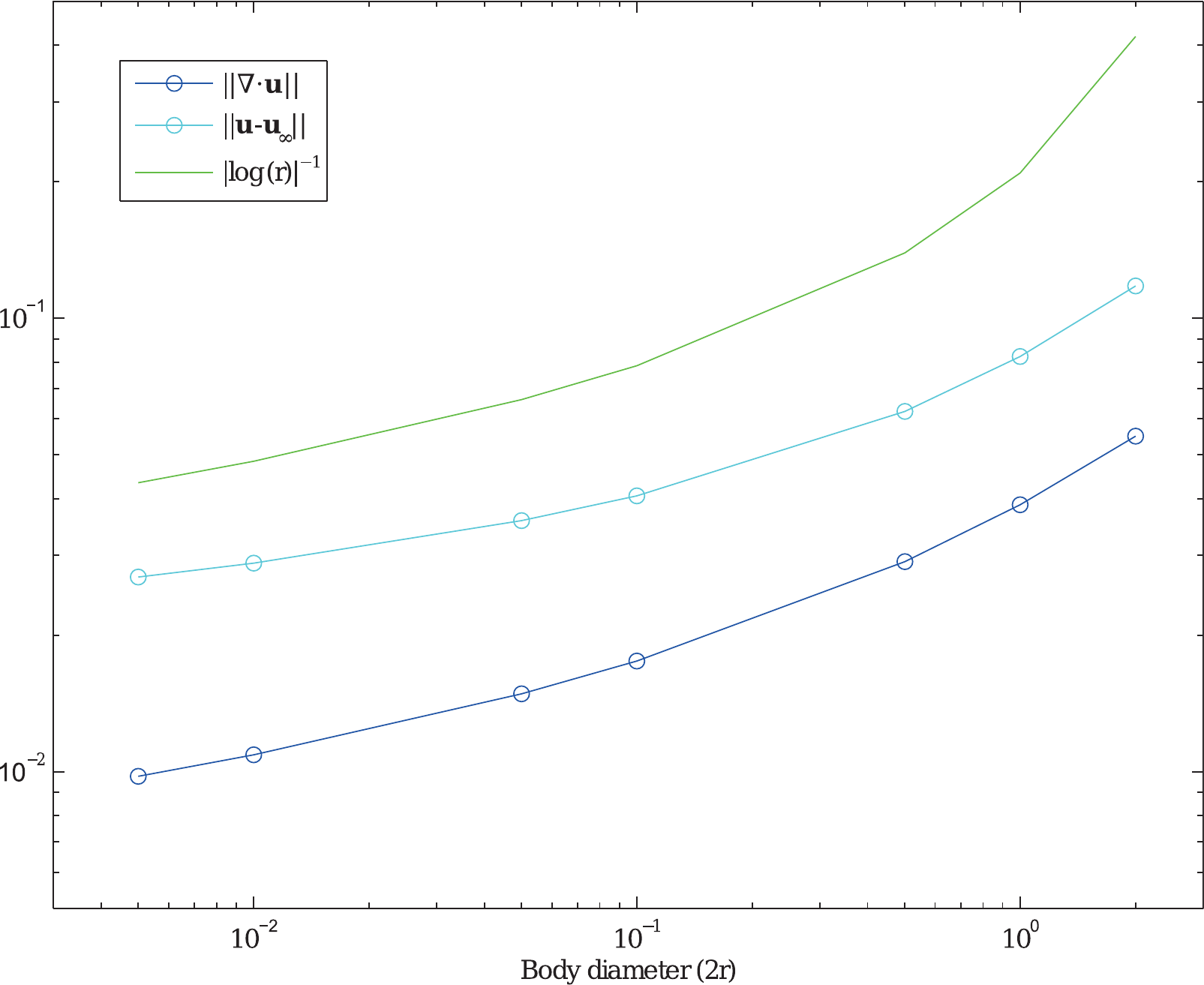}
 \end{center}
 \caption{\label{fig:gradunorm}The norms of the velocity gradient and the velocity field as seen by the body, as a function of body size normalized by the fluid domain surface.}
\end{figure}

 Figure~\ref{fig:gradunorm} shows the evolution of these norms as a function of body diameter ($2r$). Over the range of body sizes that we investigated, the results suggest that the norm of the velocity gradient vanishes as $\vert\log{(r)}\vert^{-1}$ as $r\to 0$.

\par This investigation shows that given the right choice of parameters, the \adaptivebc\ may be used to simulate flows with smaller Reynolds number (in this run $\ReN = 2r$) than what was expected at first from the results of the validation. Indeed, here the body size is decreased and the viscous length remains the same (resulting in a decrease of the Reynolds number), and thus the \abc,\ which explicitly depend on $\lv$, are always evaluated at some well-adapted distance regardless of body size, whose effect is only reflected through the value of $c_1^ \ast$. This is in contrast to the simulations presented in Section~\ref{sec:val}, where the body size was kept fixed, and the Reynolds number (and thus the viscous length) changed, which would have required to adapt the domain size appropriately for the \abc\ to be useful.


\section{\label{sec:forcewall}Force as a function of wall distance}

\subsection{Mesh}

\par In this investigation, we keep to the triangle-only family of meshes, similar to the ones already used in Sections~\ref{sec:shape} and \ref{sec:theobehavior}. We choose a domain truncation length $l=40$ for all simulations, and the most refined meshes are composed of almost 370'000 elements and approaching 1.7 million degrees of freedom.

\subsection{Results}

\par
\begin{figure}[!h]
 \begin{center}
  \includegraphics[width=\textwidth]{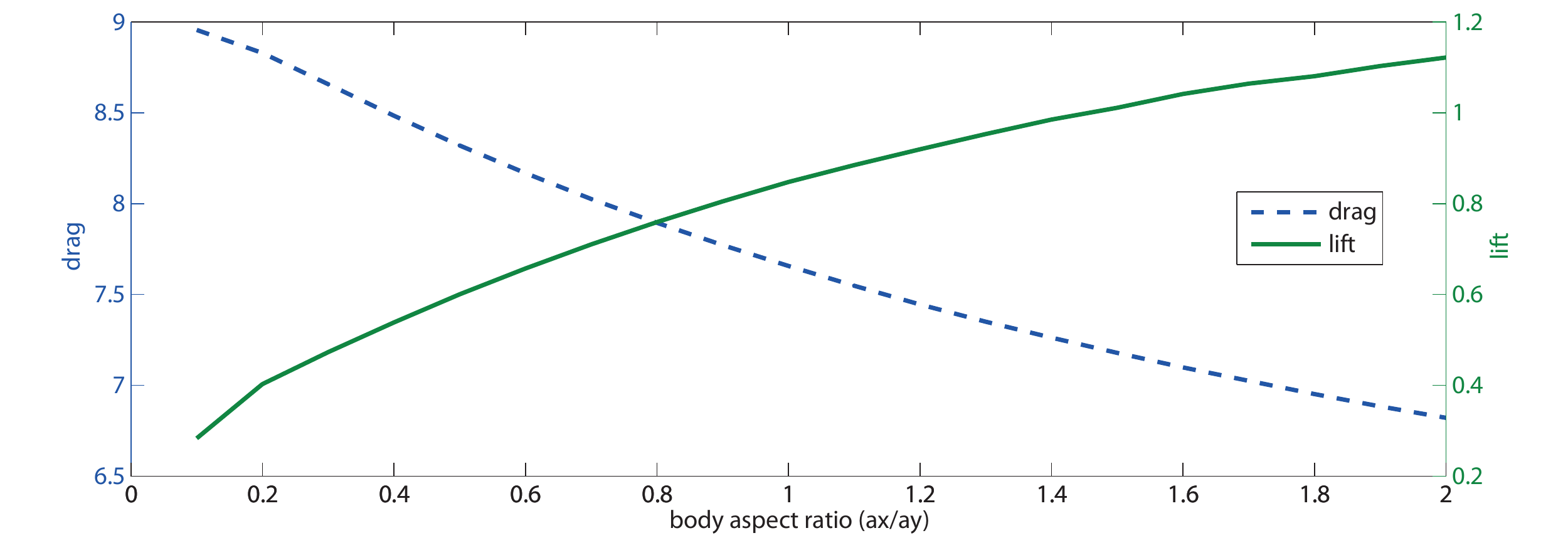}
 \end{center}
 \caption{\label{fig:ellAR}Hydrodynamic forces at $\ReN=1$ versus ellipse aspect ratio, for a body to wall distance $d=1$.}
\end{figure}

 We only treat the case of bodies with slip boundary condition, because, according to \cite{Takemura.Magnaudet-transverseforceclean2003}, it is this type of body which is susceptible to experience a zero transverse force for a certain body-wall distance. We present the results for simulations for bodies with circular and elliptical shapes. For the case of the elliptic bodies, we choose a low ratio of axes $a_x/a_y$, since this is the shape with the smallest lift for a given body-wall distance (see Figure~\ref{fig:ellAR}). An ellipse with its longer axis perpendicular to the flow direct may also act as a simple model for bubbles in flows of Reynolds number $\ReN \sim \mathcal{O}(100)$, where real bubbles flatten, see \cite{Zenit.Magnaudet-Pathinstabilityof2008}.
\par
\begin{figure}[!h]
 \begin{center}
  \includegraphics[width=\textwidth]{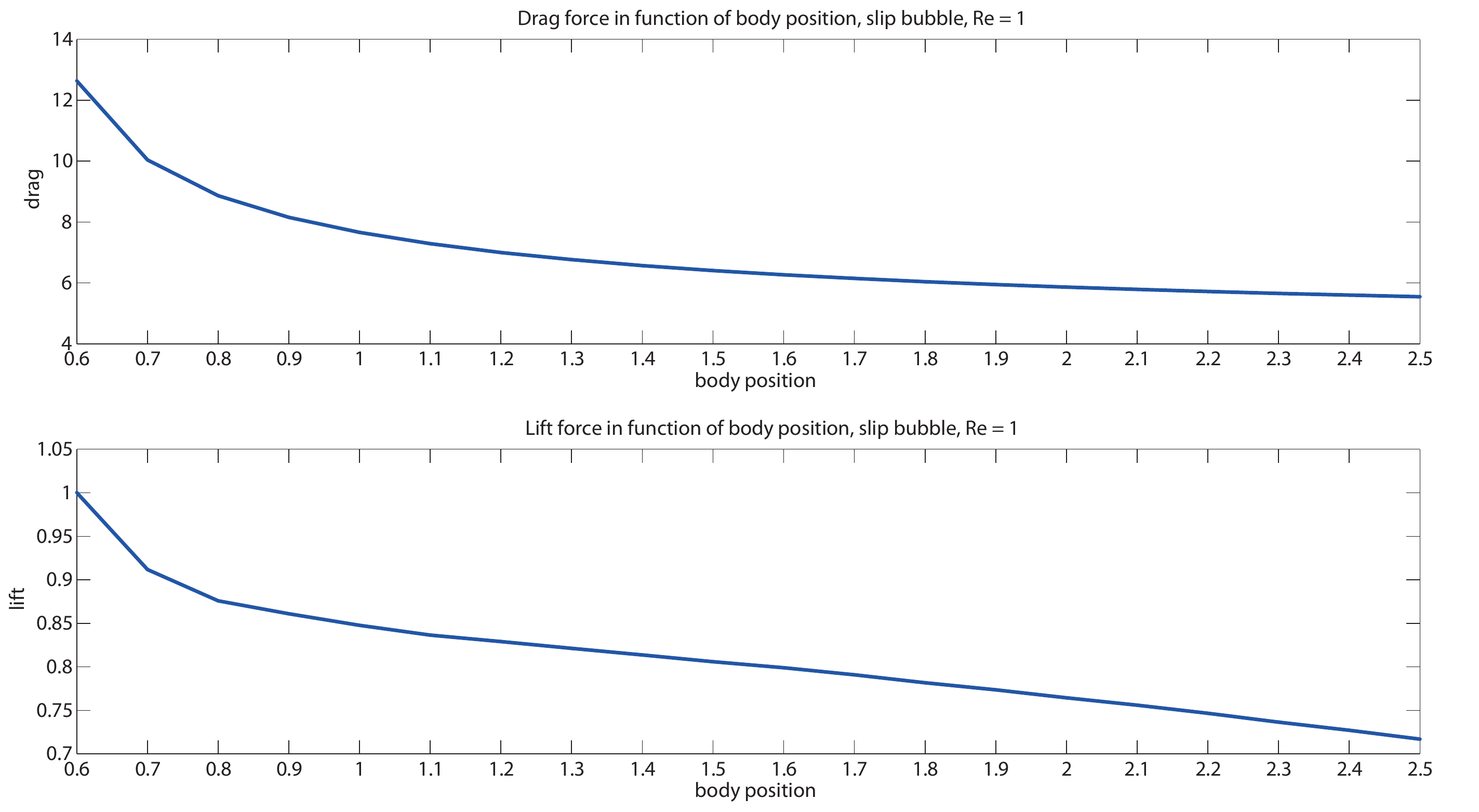}
 \end{center}
 \caption{\label{fig:forcewallCircRe1}Hydrodynamic forces versus body-wall distance, for a circular object, at $\ReN=1$.}
\end{figure}

\begin{figure}[!h]
 \begin{center}
  \includegraphics[width=\textwidth]{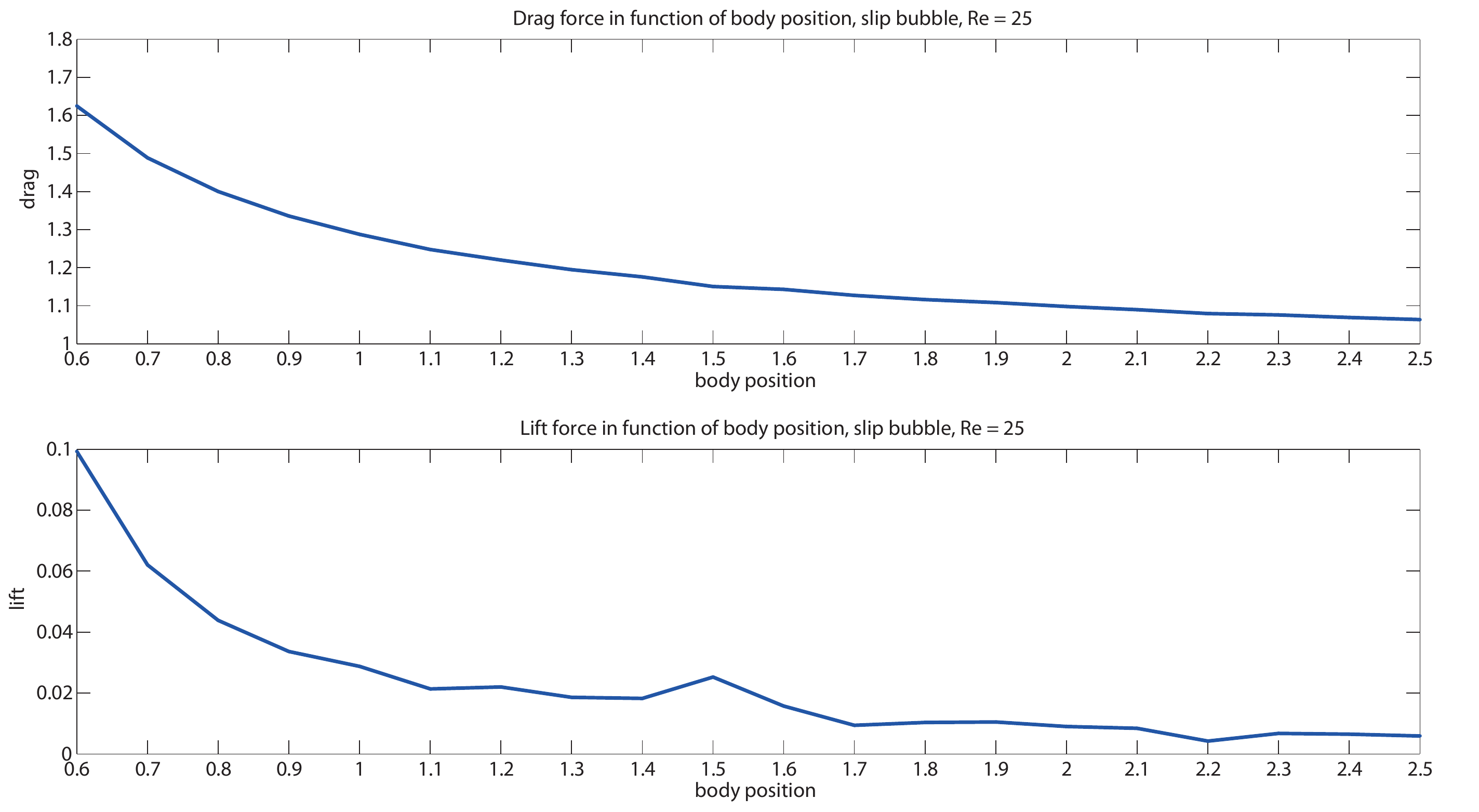}
 \end{center}
 \caption{\label{fig:forcewallEll01Re25}Hydrodynamic forces versus body-wall distance, for an elliptical object with aspect ratio $a_x/a_y=0.1$, at $\ReN=25$.}
\end{figure}

 Again we tested the range of Reynolds numbers $\ReN = 0.5,1,5,10,25$ (obtained numerically by changing the dynamic viscosity $\nu$), for the range $d=0.6,\ldots,2.5$ (with increments in steps of $0.1$). Figure~\ref{fig:forcewallCircRe1} shows the hydrodynamic forces in the case of a circular body at $\ReN = 1$, and Figure~\ref{fig:forcewallEll01Re25} shows the same forces for the elliptic body with $a_x/a_y = 0.1$ at $\ReN = 25$. In none of these combinations of parameters do we observe a change of sign in the lift, in contrast to what is predicted by the experiments for the three dimensional case.

\section{Conclusion}

\par We have presented simulations of flow around bodies in a half-plane for a wide range of parameters, including a range of three orders of magnitude of Reynolds numbers for some cases. We have shown that thanks to the \adaptivebc, the dependence of the computed values for drag and lift on the size of the computational domain is drastically reduced, achieving accuracy better by one to two orders of magnitude when compared to simulations with simple or classic boundary conditions. Therefore, with the \abc\ a given accuracy can be obtained on much smaller domains, thus bringing down the hardware requirements (CFD on a laptop). We have also shown a substantial qualitative improvement of the physical behavior of the flow, in the sense that the \adaptivebc\ have a minimal influence on the streamlines, in particular close to the artificial boundaries.
\par The results obtained in the numerical simulations seem to indicate that the lift monotonically decreases with the distance to the wall, but no point where the transverse force vanishes could be found. It is an open question if this experimental result depends on dimensionality, on the exact geometry of the body or fluid container, or if it is  the result of the flow being unsteady.
\par Finally, the \adaptivebc\ has been shown to work well for a variety of smooth bodies, even in collections.

\section*{Acknowledgments}

We would like to thank Julien Guillod and Matthieu Hillairet for fruitful discussions on the topic of this article and related matters.


\section{References}

\bibliographystyle{amsplain}
\bibliography{../../papers/CompleteDatabase2011}

\newpage


\appendix

\section{\label{sec:explicitasfun}Deriving the asymptotic boundary condition}

We show how to derive the \adaptivebc\ (\ref{eq:asstructu}) and (\ref{eq:asstructv}). In \cite{Hillairet.Wittwer-Asymptoticdescriptionof2011} it was shown that the non-dimensional system 
\begin{align}
\partial _{\shifted{x}}\shifted{\boldsymbol{u}} + \shifted{\boldsymbol{u}}\cdot \shifted{\mathbf{\nabla}}\shifted{\boldsymbol{u}} + \shifted{\mathbf{\nabla}}\shifted{p} - \shifted{\Delta} \shifted{\boldsymbol{u}} & = 0~, \label{eq:nsAsPaper} \\
\shifted{\mathbf{\nabla}} \cdot \shifted{\boldsymbol{u}} & = 0~,  \label{eq:incAsPaper}
\end{align}%
in the domain $\shifted{\Omega}=\{(\shifted{x},\shifted{y})\in \mathbb{R}\times[1,\infty)\setminus \bubble\},$ is equivalent to the same problem without the body, but with some force term of compact support, in the sense that the solutions coincide outside some compact set containing the body. For the system without a body, an asymptotic expansion of the velocity field (with $\shifted{y}^{-1}$ playing the role of the small parameter) was obtained in \cite{Boeckle.Wittwer-Asymptoticsofsolutions2011}, given by
\begin{align}
\shifted{u}_{\mathrm{as}}(\shifted{x},\shifted{y})& = \frac{c_1}{\shifted{y}^{3/2}} \varphi_1(\shifted{x}/\shifted{y}) + \frac{c_1}{\shifted{y}^2} \varphi_{2,1}(\shifted{x}/\shifted{y}) + \frac{c_2}{\shifted{y}^2} \varphi_{2,2}(\shifted{x}/\shifted{y}) \notag \\
 & - \frac{c_1}{\shifted{y}^2} \eta_W(\shifted{x}/\shifted{y}^2) - \frac{c_1}{\shifted{y}^3} \eta_B(\shifted{x}/\shifted{y}^2)~, \label{eq:uasAsPaper} \\
\shifted{v}_{\mathrm{as}}(\shifted{x},\shifted{y})& = \frac{c_1}{\shifted{y}^{3/2}} \psi_1(\shifted{x}/\shifted{y}) + \frac{c_1}{\shifted{y}^2} \psi_{2,1}(\shifted{x}/\shifted{y}) + \frac{c_2}{\shifted{y}^2} \psi_{2,2}(\shifted{x}/\shifted{y}) \notag \\
 & + \frac{c_1}{\shifted{y}^3} \omega_W(\shifted{x}/\shifted{y}^2) + \frac{c_1}{\shifted{y}^4} \omega_B(\shifted{x}/\shifted{y}^2)~, \label{eq:vasAsPaper}
\end{align}
where $(\shifted{x},\shifted{y})\in \mathbb{R\times \lbrack }1,\infty)$, and where the functions $\varphi_1$, $\varphi_{2,1}$, $\varphi_{2,2}$, $\psi_1$, $\psi_{2,1}$, $\psi_{2,2}$, $\eta_B$, $\eta_W$, $\omega_B$ and $\omega_W$ are as defined in (\ref{eq:asphi1})--(\ref{eq:aspsi22}) and (\ref{eq:etaW})--(\ref{eq:omegaB}). The constants $c_1$ and $c_2$ depend on the solution and are defined more precisely in \cite{Boeckle.Wittwer-Asymptoticsofsolutions2011}. The equations (\ref{eq:nssteady}) and (\ref{eq:incompressibility}) are obtained from (\ref{eq:nsAsPaper}) and (\ref{eq:incAsPaper}) by setting
\begin{align}
  \shifted{\boldsymbol{u}}(\shifted{\boldsymbol{x}}) & = \vinfM^{-1}\boldsymbol{u}(\boldsymbol{x}) - \boldsymbol{e}_1 ~, \label{eq:varchange}\\
  \tabvec{\shifted{x}}{\shifted{y}} & =  \tabvec{\lv^{-1} x}{\lv^{-1} y + 1}~. \label{eq:coordchange}
\end{align}
Inserting this into (\ref{eq:uasAsPaper}) and (\ref{eq:vasAsPaper}) we get%
\begin{align}
\vinfM^{-1} u_{\mathrm{as}}(x,y) & = 1 + \shifted{u}_{\mathrm{as}}(x/\lv,1+y/\lv)) \nonumber \\
& =  1 + \frac{c_1}{(1+y/\lv)^{3/2}}\varphi_1(\shifted{x}/\shifted{y}) \nonumber \\
& + \frac{c_1}{(1+y/\lv)^{2}}\varphi_{2,1}(\shifted{x}/\shifted{y}) + \frac{c_2}{(1+y/\lv)^{2}}\varphi_{2,2}(\shifted{x}/\shifted{y})\nonumber \\
& - \frac{c_1}{(1+y/\lv)^{2}}\eta_W(\shifted{x}/\shifted{y}^{2}) - \frac{c_1}{(1+y/\lv)^{3}}\eta_B(\shifted{x}/\shifted{y}^{2})~. \label{eq:expandinguas}
\end{align}
For $y\to\infty$ we have, for example
\begin{align*}
 \frac{1}{(1+y/\lv)^2} \sim \frac{\lv^2}{y^2} -\frac{2\lv^3}{y^3} + \ldots
\end{align*}
and
\begin{align*}
 \eta_W \left( \frac{x/\lv} {\left( 1 + y/\lv \right) ^2} \right) & = \eta_W \left( \frac{x/\lv}{(y/\lv)^2} \left(1+(\lv/y)\right)^{-2}\right) \\
 & \sim \eta_W \left( \frac{x/\lv}{(y/\lv)^2} - \frac{2\lv}{y}\frac{x/\lv}{(y/\lv)^2}+ \ldots \right) \\
 & \sim \eta_W \left( \frac{\lv x}{y^2} \right) - \frac{2\lv}{y}\frac{\lv x}{y^2} \eta_W^{\prime}\left( \frac{\lv x}{y^2} \right) + \ldots
\end{align*}
where we have used the Taylor series of the function $\eta_W$ for small $\lv x/y^2$. Applying this to each term in (\ref{eq:expandinguas}), then discarding terms decaying faster than $y^{-2}$ obtained from the $\shifted{x}/\shifted{y}$ scale, and decaying faster than $y^{-3}$ obtained from the $\shifted{x}/\shifted{y}^2$ scale, and remembering that any explicit $x$ be grouped with scaling-appropriate powers of $y^{-1}$, we get
\begin{align*}
\vinfM^{-1} u_{\mathrm{as}}(x,y) & \sim 1 + \shifted{u}_{\mathrm{as}}(x/\lv,y/\lv) - \frac{2\lv^3c_1}{y^{3}}\eta_W(\lv x/y^{2}) - \frac{2\lv^3c_1}{y^{3}}\frac{\lv x}{y^{2}}\eta_W^{\prime}(\lv x/y^{2})~,
\end{align*}%
and similarly, this time discarding terms decaying faster than $y^{-4}$ obtained from the $\shifted{x}/\shifted{y}^2$ scale
\begin{align*}
\vinfM^{-1} v_{\mathrm{as}}(x,y) & \sim \shifted{v}_{\mathrm{as}}(x/\lv,y/\lv) - \frac{3\lv^4c_1}{y^{4}}\omega_W(\lv x/y^{2}) - \frac{2\lv^4c_1}{y^{4}}\frac{\lv x}{y^{2}}\omega_W^{\prime}(\lv x/y^{2})~.
\end{align*}%
This procedure is a resummation technique, which happens to yield a new expansion which respects the boundary condition (\ref{eq:bndmvwall}) if we set $c_2=0$, \ie
\begin{equation*}
\lim_{\substack{y\rightarrow 0 \\ x\in \mathbb{R}\backslash\{0\}}} \boldsymbol{u}_{\mathrm{as}}(x,y) = \boldsymbol{u}(x\neq 0,0) = \vinf~.
\end{equation*}
\begin{remark}\label{rem:exchangelimits}
It is the fact that the asymptotic expansions involve scale-invariant functions which makes this exchange of limits possible, allowing to extend the validity of the expansion, which was originally valid for large $y$ only, also to large $x$. A detailed mathematical proof of this feature has however not been carried out yet.
Expressions (\ref{eq:asstructu}) and (\ref{eq:asstructv}) now follows from these results, with the constants $c^\ast_1$ and $c^\ast_2$ replacing $c_1$ and $c_2$ since the domain is finite in the numerical implementation (see \ref{sec:c1algorithm} for more details).
\end{remark}


\section{\label{sec:c1algorithm}Approximating the unknown constant of the asymptotic expansion}

We motivate the choice of the function $g$ given in (\ref{eq:gdef}). First, we must find an asymptotic description for the pressure, valid in the domain $\domain$. Using the Ansatz
\begin{align*}
 p = -\frac{1}{2}\boldsymbol{u}_E\cdot\boldsymbol{u}_E + \rho~,
\end{align*}
where the index $E$ means that we only retain functions with $x/y$ as the argument in the asymptotic expansion of the velocity. Inserting this into (\ref{eq:nssteady}) we get
\begin{align*}
 \boldsymbol{u}\cdot\mathbf{\nabla}\boldsymbol{u} - \boldsymbol{u}_E\cdot\mathbf{\nabla}\boldsymbol{u}_E + \boldsymbol{\nabla}\rho- \nu \Delta\boldsymbol{u} & = 0~,
\end{align*}
and then taking the divergence yields
\begin{align*}
 \Delta \rho = - \left(\nabla \boldsymbol{u}\right)^\mathrm{T}:\left(\nabla \boldsymbol{u}\right) + \left(\nabla \boldsymbol{u}_E\right)^\mathrm{T}:\left(\nabla \boldsymbol{u}_E\right), 
\end{align*}
where \textquotedblleft~$:$~\textquotedblright\ denotes the tensorial scalar product (\ie $\mathrm{A}:\mathrm{B} = \sum_{i,j}a_{ij}b_{ij}$). The dominant terms on the r.h.s.\ are products of functions of $x/y$ and $x/y^2$, or $x/y^2$ and $x/y^2$, decaying at least as fast as $y^{-11/2}$. Since for large $y$ the dominant terms on the l.h.s.\ are those obtained from $\partial_y^2$, integration shows that $\rho \sim y^{-7/2}$, which is beyond the terms we retain, so that to leading order the pressure is given by
\begin{align}
 p & \sim -\vinfM^2\left(\frac{1}{2} + \frac{c_1}{(y/\lv)^{3/2}}\varphi_1(x/y) + \frac{c_1}{(y/\lv)^2}\varphi_{2,1}(x/y) \right). \label{eq:asstructp}
\end{align}
Second, we obtain a formula which gives the constant $c_1^\ast$ as a ratio between an integral over the domain $\truncated{\domain}$ and an integral over its boundaries. We define the tensor $\mathrm{T}$ by%
\begin{equation*}
\mathrm{T} = -\boldsymbol{u}\otimes\boldsymbol{u} + \nu\mathcal{D}(\boldsymbol{u}) - pI~, 
\end{equation*}
where $\otimes$ represents the dyadic product (\ie $(\boldsymbol{a}\otimes\boldsymbol{b})_{ij} = a_ib_j$) and $\mathcal{D}(\boldsymbol{u})= \nabla\boldsymbol{u}+(\nabla\boldsymbol{u})^{\mathrm{T}}$, with $\nabla\cdot\mathrm{T}$ yielding (\ref{eq:nssteady}). We choose the vector%
\begin{equation*}
\boldsymbol{V}=(\sqrt{y}\chi_{\truncated{\domain}}(x,y),0)^{\mathrm{T}}~, 
\end{equation*}
where the factor $\sqrt{y}$ is used for reasons that will become clear
later on and where 
\begin{equation*}
\chi_{\truncated{\domain}}(x,y)=\chi_{x}(x)\cdot\chi_{y}(y) 
\end{equation*}
is a cutoff function which cuts a channel perpendicular to the wall centered on the body of radius $r$ and two strips, one adjacent to the wall and one adjacent to the artificial boundary parallel to the wall of the domain $\truncated{\domain}=\truncated{\domain}_+\setminus\bubble$, where $\truncated {\domain}=(-l,l)\times(0,l)$ and $\bubble=\{\boldsymbol{x}\in\truncated{\domain}\mid x^{2}+(y-d)^{2}\leq r^{2}\}$ ($d$ is the distance between the body center and the wall), with $l>d+r$ and $d>r>0$. Namely,%
\begin{equation*}
\chi_{x}(x)=\chi(-(r+x)/4)\cdot\chi((x-r)/4) 
\end{equation*}
and%
\begin{equation*}
\chi_{y}(y)=\chi(y/4)\cdot\chi((l-y)/4)~, 
\end{equation*}
where $\chi$ is an arbitrary smooth cutoff function which we have chosen as follows,
\begin{align*}
\chi(\eta) & =\frac{1}{\rho}\int_{0}^{\max\{0,\min\{1,\eta\}\}}\eta(1-\eta)^{4}d\zeta~, \\
\partial_{\eta}\chi(\eta) & =\left\{ 
\begin{array}{rl}
\frac{1}{\rho}\eta(1-\eta)^{4} & ,~\eta\in[0,1] \\ 
0 & ,~\mathrm{otherwise}%
\end{array}
\right.
\end{align*}
and $\rho=\int_{0}^{1}\eta(1-\eta)^{4}d\eta$. The factor $1/4$ in the arguments of $\chi_{x}$ and $\chi_{y}$ is arbitrary and is chosen such as to have numerically reasonable gradients. The power $4$ in the definition of $\chi$ is arbitrary as well and is simply chosen to yield a sufficiently smooth expression. We have%
\begin{equation*}
\mathrm{T}\boldsymbol{V}=\sqrt{y}\chi_{\truncated{\domain}}(x,y)\left( -%
\begin{pmatrix}
u^{2} \\ 
uv%
\end{pmatrix}
+\nu%
\begin{pmatrix}
2\partial_{x}u \\ 
\partial_{x}v+\partial_{y}u%
\end{pmatrix}
-%
\begin{pmatrix}
p \\ 
0%
\end{pmatrix}
\right) ~. 
\end{equation*}
Integrating over the whole domain, we get, by Gauss's theorem,%
\begin{equation}
\int_{\truncated{\domain}}\nabla\cdot\mathrm{T}\boldsymbol{V}d\omega = \int_{\partial\bubble}(\mathrm{T}\boldsymbol{V})\cdot\boldsymbol{n}d\sigma + \int_{\partial\truncated{\domain}}(\mathrm{T}\boldsymbol{V})\cdot\boldsymbol{n}d\sigma~.  
\end{equation}
The integral over the surface of the body $\partial\bubble$ vanishes by the choice of the cutoff function. We have%
\begin{align}
I_{\truncated{\domain}}& :=\int_{\truncated{\domain}}\nabla\cdot \mathrm{T} \boldsymbol{V}d\omega \notag \\
 & = \int_{\truncated{\domain}}\chi _{x}(x)\left( \frac{\chi_{y}(y)}{2\sqrt{y}}+\sqrt{y}\partial_{y}\chi _{y}(y)\right) ( -uv + \nu\partial_x v + \nu\partial_y u ) d\omega  \notag \\
& +\int_{\truncated{\domain}} \sqrt{y}\chi_{y}(y)\partial_{x}\chi_{x}(x) ( -u^2 + 2\nu\partial_x u - p ) d\omega~, \label{eq:c1VolumeIntegral}
\end{align}
which is computed numerically from the FEM solution. Due to the choice of the cutoff function, we also have
\begin{align}
I_{\partial\truncated{\domain}}:=\int_{\partial\truncated{\domain}}(\mathrm{T}\boldsymbol{V}) \cdot \boldsymbol{n}d\sigma & = \int_{l}^{0}\left. \sqrt{y}\chi_{\truncated{\domain}}(x,y) (-u^2 + 2\nu\partial_x u-p) \right\vert_{x=-l} (-dy) \notag \\
& + \int_{0}^{l}\left. \sqrt{y}\chi_{\domain}(x,y) ( u^2 - 2\nu\partial_x u + p ) \right\vert _{x=+l}dy~. \label{eq:c1BorderIntegral}
\end{align}
On $\partial\truncated{\domain}$ we use (\ref{eq:asstructu}) and (\ref{eq:asstructv}) to represent the velocity field. We do not consider terms which decay faster than $1/y^{2}$ in the $x/y$ scaling, since these would be of the same order as those we neglect in our asymptotic expansion. In the $x/y^{2}$ scaling, we neglect those terms which decay faster than $1/y^{3}$ for the same reasons. Mixed terms (comprised of a product of functions of either scaling behavior) are neglected if they decay faster than $1/y^{2}$, meaning that none are actually retained. We obtain%
\begin{align*}
u^{2} \sim &~ \vinfM^2 \left( 1 + \frac{2c_1^\ast}{(y/\lv)^{3/2}}\varphi_1(x/y) + \frac{2c_1^\ast}{(y/\lv)^2}\varphi_{2,1}(x/y)\right) \\
 & +\vinfM^2 \left( - \frac{2c_1^\ast}{(y/\lv)^2}\eta_1(\lv x/y^2) - \frac{2c_1^\ast}{(y/\lv)^3}\eta_2(\lv x/y^2) \right), \\
-2\partial_{x}u  \sim & ~ 0~, \\
p  \sim & -\frac{1}{2}(u_E^2+v_E^2) \\
   \sim & -\vinfM^2 \left( \frac{1}{2} + \frac{c_1^\ast}{(y/\lv)^{3/2}}\varphi_1(x/y) + \frac{c_1^\ast}{(y/\lv)^{2}}\varphi_{2,1}(x/y)\right),
\end{align*}
so that
\begin{align*}
& \left. (u^{2}-2\partial_{x}u+p)\right\vert _{x=+l}-\left.
(u^{2}-2\partial_{x}u+p)\right\vert _{x=-l} \\
& \sim \frac{c_1^\ast\lv^{3/2}}{y^{3/2}}\left(\varphi_1(l/y)-\varphi_1(-l/y)\vphantom{y^2}\right) + \frac{c_1^\ast\lv^2}{y^2}\left(\varphi_{2,1}(l/y)-\varphi_{2,1}(-l/y)\vphantom{y^2}\right) \\
& -\frac{2c_1^\ast\lv^2}{y^2}\eta_1(\lv l/y^2)+0 - \frac{2c_1^\ast\lv^3}{y^3}\left(\eta_2(\lv l/y^2)-\eta_2(-\lv l/y^2)\right) \\
& \sim \frac{c_1^\ast\lv^{3/2}}{y^{3/2}}\left(\varphi_1(l/y)-\psi_1(l/y)\vphantom{y^2}\right) + \frac{2c_1^\ast\lv^2}{y^2}\varphi_{2,1}(l/y) \\
& -\frac{2c_1^\ast\lv^2}{y^2}\eta_1(\lv l/y^2) - \frac{2c_1^\ast\lv^3}{y^3}\left(\eta_2(\lv l/y^2)-\eta_2(-\lv l/y^2)\right)~,
\end{align*}
where we use that $\varphi_{1}(-z)=\psi_{1}(z)$ and that $\varphi_{2,1}$ is an odd function. We insert this approximation into (\ref{eq:c1BorderIntegral}), where $\chi_{\truncated{\domain}}(\pm l,y)=\chi_{y}(y)$ (\ie the fact that the computational domain is of equal length upstream and downstream simplifies the expressions), and treat the integral separately according to the two scalings. We get%
\begin{align*}
I_{\partial\truncated{\domain}}^{x/y} & := c_1^\ast \lv^{3/2}\int_0^l \chi_y(y)\frac{\varphi_1(l/y)-\psi_1(l/y)}{y}dy + 2c_1^\ast \lv^2 \int_0^l \chi_y(y)\frac{1}{y^{3/2}}\varphi_{2,1}(l/y)dy \\
& \overset{z=l/y}{=} c_1^\ast \int_1^{\infty}\chi_y(l/z)\left(\lv^{3/2} \frac{\varphi_1(z)-\psi_1(z)}{z} + 2\lv^2 \frac{\varphi_{2,1}(z)}{\sqrt{zl}}\right)dz~,
\end{align*}
where we see that the factor $\sqrt{y}$ in $\boldsymbol{V}$ is chosen in order to obtain a non-vanishing expression for large $l$, and%
\begin{align*}
I_{\partial\truncated{\domain}}^{x/y^2} & := -2 c_1^\ast \lv^2\int_0^l \chi_y(y)\frac{\eta_1(\lv l/y^2)}{y^{3/2}}dy - 2 c_1^\ast \lv^3 \int_0^l \chi_y(y)\frac{\eta_2(\lv l/y^2)-\eta_2(-\lv l/y^2)}{y^{5/2}}dy \\
& \overset{z=\lv l/y^2}{=} - c_1^\ast \int_{\lv/l}^{\infty}\chi_y(\sqrt{\lv l/z})\left( \lv^2\frac{\eta_1(z)}{(\lv lz^3)^{1/4}} + \lv^3\frac{\eta_2(z)-\eta_2(-z)}{\left(\lv^3 l^3 z\right)^{1/4}}\right)dz~.
\end{align*}
Note that in the derivation of the integrals on the upstream and downstream boundary of the truncated domain $\truncated{\domain}$, we assumed that the asymptotic expansion may be extended, for large fixed $x$, to all $y$ although the asymptotic expansion is a priori only valid for large $y$. See Remark~\ref{rem:exchangelimits} in \ref{sec:explicitasfun} for a motivation.

We then define%
\begin{equation*}
  n_1(l,\lv,\chi_{\truncated{\domain}}):=(I_{\partial\truncated{\domain}}^{x/y}+I_{\partial\truncated {\domain}}^{x/y^2})/c_1^\ast~, 
\end{equation*}
which can be computed numerically. We then have that
\begin{equation*}
  c_1^\ast=c_1^\ast(l)=\frac{I_{\truncated{\domain}}}{n_1}~,
\end{equation*}
and we expect that the constant $c_1$ associated to the solution of the original problem is given by
\begin{equation*}
  c_1=\lim_{l\to\infty}c_1^\ast(l)~.
\end{equation*}
This motivates the definition of (\ref{eq:gdef}).

\end{document}